\documentclass[a4paper]{jpconf}

\usepackage{tikz}

\usepackage[pass]{geometry} 
\usepackage{color, braket,bm,subfig, mathrsfs,mathtools, wasysym,amsmath, amsfonts,amssymb,graphicx, slashed, cite}

\newcommand{\col}{\textcolor{blue}}

  \DeclareMathAlphabet{\mathpzc}{OT1}{pzc}{m}{it}

\makeatletter

\def\ps@headings{%
     \def\@oddfoot{\hfil\thepage\hfil}
     \def\@evenfoot{\hfil\thepage\hfil}
     \let\@oddhead\@empty
     \let\@evenhead\@empty
      \let\@mkboth\markboth
      \let\sectionmark\@gobble
      \let\subsectionmark\@gobble}

\makeatother

\begin{document}

%%%%%%%%%%%%%%%%%% remove for submission %%%%%%%%%%%%%%%%

%\begin{flushright}
%\begin{footnotesize}
% MAN/HEP/2015/04\\
% March 2015
%\end{footnotesize} 
%\end{flushright}

%%%%%%%%%%%%%%%%%%%%%%%%%%%%%%%%%%%%%%%%%%%%%%%%%%%%%%%%%

\title{Natural Alignment in the Two Higgs Doublet Model}

\author{P.~S.~Bhupal Dev$^1$, \underline{Apostolos Pilaftsis}$^2$}

\address{$^1$Department of Physics and McDonnell Center for the Space Sciences,  Washington University, St. Louis, MO 63130, USA}
\address{$^2$Consortium for Fundamental Physics,
  School of Physics and Astronomy, University of Manchester, Manchester M13 9PL, United Kingdom.}
\ead{bdev@wustl.edu, apostolos.pilaftsis@manchester.ac.uk}
%%%%%%%%%%%%%%%%%%%%%%%%%%%%%%%%%%%%%%%%%%%%%%%%%%%%%%%%%%%%%%%%%%%%%%%%%%%
\begin{abstract}
As the LHC Higgs data persistently suggest the couplings
       of the observed 125 GeV Higgs boson to be consistent with the Standard
       Model (SM) expectations, any extended Higgs sector must lead to
       the so-called SM {\em alignment limit}, where one of the Higgs
       bosons behaves exactly like that of the SM. In the context of
       the Two Higgs Doublet Model (2HDM), this alignment is often
       associated with either decoupling of the heavy Higgs sector or
       accidental cancellations in the 2HDM potential. We present a
       novel symmetry justification for `natural' alignment without
       necessarily decoupling or fine-tuning. We show that there exist
       only {\em three} different symmetry realizations of the natural
       alignment scenario in 2HDM. We identify the 2HDM parameter space satisfying the natural alignment condition up to the Planck scale. We also analyze new collider signals for
       the heavy Higgs sector in the natural alignment limit, which
       dominantly lead to third-generation quarks in the final state
       and can serve as a useful observational tool during the Run-II
       phase of the LHC.
\end{abstract}
%%%%%%%%%%%%%%%%%%%%%%%%%%%%%%%%%%%%%%%%%%%%%%%%%%%%%%%%%%%%%%%%%%%%%%%%%%%%%
\section{Introduction}\label{sec:1}
The discovery of a Higgs boson in the Run-I phase of the LHC~\cite{Aad:2012tfa} provides the first experimental evidence for the Higgs mechanism~\cite{Higgs:1964pj} as the standard theory of electroweak symmetry breaking (EWSB). As more data are being collected at the LHC, the  coupling measurements of  the  observed Higgs  boson seem to be very close to the  Standard Model (SM) predictions~\cite{Khachatryan:2016vau}. The constraints deduced from the Higgs signal strength data severely limit the form of a possible heavy scalar sector in the observable sub-TeV range,  as predicted by various well-motivated  new-physics scenarios, such as supersymmetry.

Here we consider the Two Higgs Doublet Model~(2HDM)~\cite{review}, where the SM Higgs doublet is supplemented by another isodoublet with hypercharge $Y=1$. 
%This model can provide new sources of spontaneous~\cite{Lee:1973iz} or explicit~\cite{Georgi:1978xz} CP violation, viable DM candidates~\cite{Silveira:1985rk} and a strong first order phase transition for electroweak baryogenesis~\cite{Kuzmin:1985mm}.   
In the doublet field space $\Phi_{1,2}$, where $\Phi_i = (\phi_i^+, \phi_i^0)^{\sf T}$, the  general 2HDM  potential reads 
\begin{align}
  \label{pot}
V \ = \ & -\mu_1^2(\Phi_1^\dag \Phi_1) -\mu_2^2 (\Phi_2^\dag \Phi_2) 
-\left[m_{12}^2 (\Phi_1^\dag \Phi_2)+{\rm H.c.}\right]  \nonumber \\ &
+\lambda_1(\Phi_1^\dag \Phi_1)^2+\lambda_2(\Phi_2^\dag \Phi_2)^2 
 +\lambda_3(\Phi_1^\dag \Phi_1)(\Phi_2^\dag \Phi_2) 
 +\lambda_4(\Phi_1^\dag \Phi_2)(\Phi_2^\dag \Phi_1) 
\nonumber \\
& 
+\left[\frac{\lambda_5}{2}(\Phi_1^\dag \Phi_2)^2%\right.\\ 
%&&\left. 
+\lambda_6(\Phi_1^\dag \Phi_1)(\Phi_1^\dag \Phi_2) 
+\lambda_7(\Phi_1^\dag \Phi_2)(\Phi_2^\dag \Phi_2)+{\rm H.c.}\right],
\end{align}
which contains  {\em four} real mass  parameters $\mu_{1,2}^2$, Re$(m^2_{12})$,
Im$(m^2_{12})$,  and {\em ten}  real  quartic couplings  $\lambda_{1,2,3,4}$,
Re($\lambda_{5,6,7})$,  and Im($\lambda_{5,6,7}$). Thus,
the   vacuum   structure   of   the   general  2HDM   can   be   quite
rich~\cite{Battye:2011jj}, as compared to the SM. 
% and in  principle,  can allow  for a  wide
%range  of  parameter space  still  compatible  with  the existing  LHC
%constraints.    

The quark-sector Yukawa Lagrangian in the general 2HDM is given by 
\begin{align}
-{\cal L}^q_Y \  = \  \bar{Q}_L(h_1^d \Phi_1+ h_2^d\Phi_2)d_R \: +
\: \bar{Q}_L(h_1^u \widetilde{\Phi}_1+ h_2^u \widetilde{\Phi}_2)u_R \; ,
\label{yuk}
\end{align}
where $\widetilde{\Phi}_i={\rm i}\sigma^2\Phi_i^*$ ($\sigma^2$ being the second Pauli matrix) are the isospin conjugates of $\Phi_i$, $Q_L=(u_L,d_L)^{\sf T}$ are the $SU(2)_L$ quark doublets and $u_R,d_R$ are 
right-handed quark singlets. To avoid potentially large 
flavor-changing neutral current processes at the tree level induced by the Yukawa interactions in~\eqref{yuk}, one imposes a discrete Z$_2$ symmetry~\cite{review} under which 
\begin{align}
\Phi_1 \to -\Phi_1,\quad \Phi_2\to \Phi_2, \quad u_{Ra}\to u_{Ra},\quad 
d_{Ra}\to d_{Ra}~~{\rm or}~~ d_{Ra}\to -d_{Ra} \; ,
\label{discrete}
\end{align}
($a=1,2,3$ being the quark family index) so that only $\Phi_2$ gives mass to up-type quarks, and 
only $\Phi_1$ or only $\Phi_2$ gives mass to down-type quarks. 
%In this case, the scalar boson 
%couplings to quarks are proportional to the quark mass matrix, as in the SM, and therefore, there is no tree-level FCNC process. 
The ${\rm Z}_2$ symmetry~\eqref{discrete} is satisfied  by four  discrete  
choices of  tree-level Yukawa  couplings between the Higgs doublets and  SM fermions, which are  known as the Type I, II, X (lepton-specific) and Y (flipped) 2HDMs~\cite{review}. Global fits to the LHC Higgs data (see e.g.,~\cite{fit1, Chowdhury:2015yja}) suggest that all four types of 2HDMs with natural flavor conservation are constrained to
lie close to the so-called SM {\it alignment limit}~\cite{Georgi:1978ri, Gunion:2002zf, Carena:2013ooa, Dev:2014yca, Das:2015mwa}, where the mass
eigenbasis of the  CP-even scalar sector aligns with  the SM gauge
eigenbasis.  
% Specifically,  in  the  Type-II  (MSSM-type)  2HDM,  the
%coupling of the  SM-like Higgs to vector bosons  is constrained to lie
%within  10\%   of  the   SM  value  at   95\%  CL~\cite{Eberhardt:2013uba}.

The SM  alignment is  often associated with  the decoupling  limit, in
which all the non-standard Higgs bosons are assumed to be much heavier
than the electroweak scale so that the lightest CP-even scalar behaves
just  like the  SM  Higgs  boson.  The  alignment  limit  can also  be
achieved,  without decoupling~\cite{Chankowski:2000an,  Gunion:2002zf,
  Carena:2013ooa}, but  for small $\tan\beta$ values,  this is usually
attributed  to accidental  cancellations  in the  2HDM potential.   We
present a novel  symmetry argument to naturally  justify the alignment
limit~\cite{Dev:2014yca}, independently of the kinematic parameters of
the theory, such as the heavy Higgs mass and~$\tan\beta$. In particular, we show 
that there exist  only {\em  three} possible  symmetry realizations  of the
scalar  potential  which  predict natural  alignment.  
%We  explicitly
%analyze  the  simplest  case,  namely, the  Maximally  Symmetric  2HDM
%(MS-2HDM), which realizes a SO(5) symmetry in the bilinear field basis
%to be discussed  in Section~\ref{sec:3}. We show that  the renormalization group (RG)  effects  due   to  the  hypercharge  gauge   coupling  $g'$  and
%third-generation  Yukawa  couplings,  as well  as  soft-breaking  mass
%parameter $m^2_{12}$, induce relevant deviations from the SO(5) limit,
%which  lead to  distinct predictions  for  the Higgs  spectrum of  the
%MS-2HDM (see Section~\ref{sec:4}).  In particular, 
A striking outcome of this analysis is that
the  heavy Higgs  sector is  predicted to  be {\em  quasi-degenerate},
apart from being {\em gaugephobic}, which  is a generic feature in the
alignment limit.  Moreover, the current experimental constraints force
the heavy  Higgs sector to  lie above  the top-quark threshold. Thus, the  dominant collider signal for  this sector involves final  states with  third-generation quarks.  We make  a detailed study of  some of  these signals,  which can be  useful for  the heavy
Higgs searches in the ongoing Run-II phase of the LHC.

%The plan of this proceedings is as follows: In Section~\ref{sec:2}, we present the natural alignment condition for a generic 2HDM scalar potential. In Section~\ref{sec:3}, we list the symmetry classifications of the 2HDM potential and identify the symmetries leading to a natural alignment. In Section~\ref{sec:4}, we analyze the MS-2HDM in presence of custodial symmetry and soft breaking effects. In Section~\ref{sec:5}, we discuss some collider phenomenology of the heavy Higgs sector in the alignment limit, with particular emphasis on the heavy Higgs sector beyond the top-threshold. Our conclusions are given in Section~\ref{sec:6}.  

%%%%%%%%%%%%%%%%%%%%%%%%%%%%%%%%%%%%%%%%%%%%%%%%%%%%%%
\section{Natural Alignment Condition}\label{sec:2}
%%%%%%%%%%%%%%%%%%%%%%%%%%%%%%%%%%%%%%%%%%%%%%%%%%%%%%
For simplicity, we consider the CP-conserving 2HDM, but our results can be easily generalized to the CP-violating case. 
%We start with the linear decomposition of the two Higgs doublets in terms of eight real scalar fields:
%\begin{eqnarray}
%\Phi_j \ = \ \left(\begin{array}{c} \phi_j^+ \\ \frac{1}{\sqrt
%    2}(v_j+\phi_j+ia_j) \end{array} \right)\; , 
%\label{expand-phi}
%\end{eqnarray}
%where $v_{1,2}$ are the vacuum expectation values (VEVs) and $v=\sqrt{v_1^2+v_2^2}=246.2$~GeV  is the SM  electroweak VEV.  After  spontaneous symmetry breaking, there are three
%Goldstone   modes  ($G^\pm,G^0$),   which   become  the   longitudinal
%components  of the $W^\pm$ and  $Z$ bosons.  Thus, 
After EWSB by the vacuum expectation values (VEVs) $v_{1,2}$ of the two scalar fields $\Phi_{1,2}$, there  are five physical scalar mass eigenstates: two CP-even ($h,H$), one
CP-odd ($a$)  and two charged  ($h^\pm$) scalars. The corresponding mass     eigenvalues     are     given by~\cite{Haber:1993an}
\begin{subequations}
\begin{align}
M^2_{h^\pm} \ & = \ \frac{m_{12}^2}{s_\beta
  c_\beta}-\frac{v^2}{2}\left( \lambda_4+ \lambda_5\right)
 +\frac{v^2}{2s_\beta c_\beta}\left( \lambda_6
c_\beta^2+ \lambda_7 s_\beta^2\right), \qquad %\label{mass0}\\
%\nonumber \\ 
M_a^2 \   = \  M^2_{h^\pm}+\frac{v^2}{2}\left( \lambda_4 - 
\lambda_5\right), \label{mass1} \\ %\nonumber\\ 
M_H^2 \ & = \ \frac{1}{2}\left[(A+B)-\sqrt{(A-B)^2+4C^2}\right], \qquad %\label{mass2}\\ %\nonumber\\
M_h^2 \  = \ \frac{1}{2}\left[(A+B)+\sqrt{(A-B)^2+4C^2}\right], \label{mass3}
\end{align} 
\end{subequations}
where we have used the short-hand notations $c_\beta\equiv \cos\beta$ and $s_\beta\equiv \sin\beta$ with $\tan\beta=v_2/v_1$ and    
\begin{subequations}
\begin{align}
A \ & = \ M_a^2s_\beta^2+v^2\left( 2\lambda_1 c_\beta^2+ 
\lambda_5 s^2_\beta+2 \lambda_6 s_\beta
  c_\beta\right), \\ %\nonumber \\ 
B \ & = \ M_a^2c_\beta^2+v^2\left( 2\lambda_2 s_\beta^2+ 
\lambda_5 c^2_\beta + 2\lambda_7 s_\beta c_\beta
  \right) , \\ 
C \ & = \ -M_a^2 s_\beta c_\beta + v^2\left( \lambda_{34}s_\beta c_\beta
  + \lambda_6 c^2_\beta + \lambda_7 s^2_\beta
  \right) . 
\end{align}
\end{subequations}
with $\lambda_{34}=\lambda_3+\lambda_4$. The mixing between the mass eigenstates in the
CP-odd  and charged  sectors is  governed by the angle $\beta$, whereas in the CP-even sector, it is governed by the angle $\alpha=(1/2)\tan^{-1}[2C/(A-B)]$. The SM Higgs field can be identified as the linear combination  
\begin{eqnarray}
H_{\rm SM} \ = \  %\phi_1 \cos\beta + \phi_2 \sin \beta \ = \ 
H\cos(\beta-\alpha)+h\sin(\beta-\alpha) \; . 
\label{HSM}
\end{eqnarray}
Thus, the  couplings of  $h$ and $H$  to the SM gauge bosons
$V=W^\pm, Z$ with  respect to the SM Higgs  couplings $g_{H_{\rm SM}VV}$ will be respectively $\sin{(\beta-\alpha)}$ and $\cos{(\beta-\alpha)}$.  
%\begin{eqnarray}
%g_{hVV} \ = \ \sin{(\beta-\alpha)} \; , \qquad g_{HVV} \ = \ \cos{(\beta-\alpha)} \; . 
%\label{coup1}
%\end{eqnarray}
The {\em SM alignment limit} is defined as $\alpha\to \beta$ (or $\alpha\to \beta-\pi/2$) when $H$ ($h$) couples to vector bosons exactly as in the SM, whereas $h$ ($H$) becomes {\em gaugephobic}.  For concreteness, we will take the alignment limit as $\alpha\to \beta$.

To derive the alignment condition, we rewrite the CP-even scalar mass matrix as
%\begin{widetext}
\begin{align}
M^2_{S} \  & = \ \left(\begin{array}{cc} A& C \\ C & B \end{array}\right) 
\  = \ \left(\begin{array}{cc}
c_\beta & -s_\beta \\ 
s_\beta & c_\beta 
\end{array}\right) \left(\begin{array}{cc}
\widehat{A} & \widehat{C} \\
\widehat{C} & \widehat{B}
\end{array}\right)
\left(\begin{array}{cc}
c_\beta & s_\beta \\ 
-s_\beta & c_\beta 
\end{array}\right)\; , \label{align2} \\
%\end{eqnarray}
%\begin{subequations}
%\begin{align}
{\rm where} \quad \widehat{A} & \ = \  2v^2 \Big[ c_\beta^4 \lambda_1 
+ s_\beta^2 c_\beta^2 \lambda_{345}  
+ s_\beta^4 \lambda_2\:  +\: 2 s_\beta c_\beta \Big( c^2_\beta \lambda_6 +
s^2_\beta \lambda_7\Big)\Big]\; , \nonumber \\
\widehat{B} & \ = \  M_a^2\: +\: \lambda_5 v^2\: +\: 2 v^2 
\Big[ s^2_\beta c^2_\beta
  \Big(\lambda_1+\lambda_2-\lambda_{345}\Big)\:
-\: s_\beta c_\beta \Big(c^2_\beta - s^2_\beta\Big) \Big(\lambda_6 
- \lambda_7\Big) \Big]\; , \label{bhat} \\
\widehat{C} & \  =   \
v^2 \Big[ s^3_\beta c_\beta \Big( 2\lambda_2-\lambda_{345}\Big) - 
c^3_\beta s_\beta \Big(2\lambda_1- \lambda_{345}\Big) + c^2_\beta
\Big( 1 - 4 s^2_\beta \Big) \lambda_6 
+ s^2_\beta \Big( 4 c^2_\beta - 1\Big) \lambda_7 \Big],  \nonumber 
%\end{subequations}
\end{align}
and  we have used the short-hand notation $\lambda_{345} \equiv  \lambda_3  +
\lambda_4    +    \lambda_5$.      
%Observe    that    $\widehat{M}^2_S$
%in~(\ref{align2}) is the respective  $2\times 2$ CP-even mass matrix
%written down in the so-called Higgs eigenbasis~\cite{Georgi, Dono, Lavoura, Botella:1994cs}.
Evidently, the SM alignment limit $\alpha \to \beta$ is obtained when 
$\widehat{C} =  0$ in~\eqref{align2}~\cite{Gunion:2002zf}. This yields the quartic equation
\begin{eqnarray}
\lambda_7 \tan^4\beta -  (2\lambda_2-\lambda_{345})\tan^3\beta + 3(\lambda_6-\lambda_7)\tan^2\beta + 
(2\lambda_1-\lambda_{345})\tan\beta - \lambda_6 \ = \ 0 \; .
\label{align-gen}
\end{eqnarray}
For {\em natural} alignment, (\ref{align-gen}) should be satisfied for {\it any} value of $\tan\beta$, which requires the coefficients of the polynomial in $\tan\beta$ to vanish identically~\cite{Dev:2014yca}.  This implies 
\begin{eqnarray}
2\lambda_1 \ = \ 2\lambda_2 \ = \ \lambda_{345}\;, \qquad \lambda_6 \ = \ \lambda_7 \ = \ 0\; . 
\label{alcond}
\end{eqnarray}
In particular, for $\lambda_6 = \lambda_7 = 0$ as in the ${\rm Z}_2$-symmetric 2HDMs, (\ref{align-gen}) has a solution 
\begin{eqnarray}
  \label{tanb}
\tan^2\beta\ =\ \frac{2\lambda_1 - \lambda_{345}}{2\lambda_2 -
\lambda_{345}}\ >\ 0 \; ,
\end{eqnarray}
independent of $M_a$. After some algebra, the simple solution (\ref{tanb}) to our {\em natural alignment condition} (\ref{align-gen}) can be shown to be equivalent to that derived in~\cite{Carena:2013ooa}.

%%%%%%%%%%%%%%%%%%%%%%%%%%%%%%%%%%%%%%%%%%%%%%%%%%%%%%%%%%%%%%%%%%%%%%%%%%%%
\section{Symmetry Classifications of the 2HDM Potential}\label{sec:3}
%%%%%%%%%%%%%%%%%%%%%%%%%%%%%%%%%%%%%%%%%%%%%%%%%%%%%%%%%%%%%%%%%%%%%%%%%%%%
%The general 2HDM potential~\eqref{pot} may exhibit three different classes of accidental symmetries. The first class of symmetries pertains to transformations of the Higgs doublets $\Phi_{1,2}$ only, but not their complex conjugates $\Phi_{1,2}^*$, and are known as the Higgs family (HF) symmetries~\cite{Ginzburg:2004vp, Ferreira:2009wh}.  The second class of symmetry transformations relates the fields $\Phi_{1,2}$ to their complex conjugates $\Phi_{1,2}^*$ and are generically termed as CP symmetries~\cite{Ferreira:2009wh}. The third class of symmetries utilize mixed HF and CP transformations that leave the $SU(2)_L$ gauge kinetic terms of $\Phi_{1,2}$ canonical~\cite{Battye:2011jj}.   
 
To identify all  accidental symmetries of the 2HDM potential~\eqref{pot},
it   is   convenient    to  work in the {\em bilinear scalar field formalism}~\cite{Maniatis:2006fs} by introducing  an  8-dimensional  complex multiplet ${\bm \Phi} \equiv (\Phi_1, \Phi_2, \widetilde{\Phi}_1, \widetilde{\Phi}_2)^{\sf T}$~\cite{Battye:2011jj,Nishi:2011gc,Pilaftsis:2011ed}. 
%\begin{equation}
%  \label{Phi}
%{\bm \Phi}\ \equiv \ \left(\begin{array}{c}  
%\Phi_1  \\
%\Phi_2  \\  
%\widetilde{\Phi}_1  \\
%\widetilde{\Phi}_2 \end{array}\right)\; ,
%\end{equation}
%where $\widetilde{\Phi}_{i} =  {\rm i}\sigma^2 \Phi^*_{i}$ (with $i=1,2$). 
%We
%should remark  that the complex  multiplet ${\bf \Phi}$  satisfies the
%Majorana property~\cite{Battye:2011jj}: ${\bf \Phi} = C {\bf \Phi}^*$,
%where  $C  =  \sigma^2  \otimes  \sigma^0  \otimes  \sigma^2$  is  the
%charge-conjugation matrix, with $\sigma^0 =
%{\bf 1}_{2\times  2}$ being the  identity matrix. 
In terms of the ${\bm \Phi}$-multiplet, we define  a {\em   null}   6-dimensional   Lorentz   vector  $R^A\ \equiv\ {\bm \Phi}^\dag \Sigma^A {\bm \Phi}$, 
%\begin{equation}
%  \label{RA}
%R^A\ \equiv\ {\bm \Phi}^\dag \Sigma^A {\bm \Phi}\; ,
%\end{equation}  
where  $A=0,1,...,5$  and  the  six $8\times  8$-dimensional  matrices
$\Sigma^A$ may be expressed in terms of the three Pauli matrices $\sigma^{1,2,3}$ and the identity matrix $\bm{1}_{2\times 2}\equiv \sigma^0$, as follows:
\begin{eqnarray}
&& \Sigma^{0,1,3} \ = \ \frac{1}{2}\sigma^0 \otimes \sigma^{0,1,3} \otimes
  \sigma^0, \qquad   
\Sigma^2 \ = \ \frac{1}{2}\sigma^3 \otimes \sigma^2 \otimes \sigma^0, \nonumber\\
&& \Sigma^4 \ = \ -\frac{1}{2}\sigma^2 \otimes \sigma^2 \otimes \sigma^0, \qquad
\Sigma^5 \ = \ -\frac{1}{2}\sigma^1 \otimes \sigma^2 \otimes \sigma^0. 
\end{eqnarray}
Note that the bilinear  field space spanned by the  6-vector $R^A$ realizes
an {\em orthochronous} ${\rm SO}(1,5)$ symmetry group~\cite{Battye:2011jj, Pilaftsis:2011ed}.

\begin{table}[t!]
\begin{center}
\begin{tabular}{c|ccccccccc}\hline
symmetry & $\mu_1^2$ & $\mu_2^2$ & $m^2_{12}$ & $\lambda_1$ & $\lambda_2$ & $\lambda_3$ & $\lambda_4$ & ${\rm Re}(\lambda_5)$ & $\lambda_6=\lambda_7$ \\ \hline
Z$_2\times $O(2) & - & - & Real & -& -& -& -& -& Real\\
$({\rm Z}_2)^2\times $SO(2) & -&  - & 0 & - & -&  - & - & - & 0 \\
$({\rm Z}_2)^3\times $O(2) & -&  $\mu_1^2$ & 0 & - &  $\lambda_1$ &  - & - & - & 0 \\
O(2)$\times $O(2) & -&  - & 0 & - & -&  - & - & 0 & 0 \\
\col{${\rm Z}_2\times $ [O(2)]$^2$} & -& $\mu_1^2$ & 0 & - & $\lambda_1$ &  - & - & $2\lambda_1-\lambda_{34}$ & 0 \\
\col{O(3)$\times $O(2)} & -&  $\mu_1^2$ & 0 & - & $\lambda_1$ &  - & $2\lambda_1-\lambda_3$  & 0 & 0 \\
SO(3) & -&  - & Real & - & - &  - & -  & $\lambda_4$ & Real \\
${\rm Z}_2\times $O(3) & - &  $\mu_1^2$ & Real & - & $\lambda_1$ &  - & -  & $\lambda_4$ & Real \\
$({\rm Z}_2)^2\times $SO(3) & -&  $\mu_1^2$ & 0 & - & $\lambda_1$ &  - & -  & $\pm \lambda_4$ & 0 \\
O(2)$\times $O(3) & -&  $\mu_1^2$ & 0 & - & $\lambda_1$ &  $2\lambda_1$ & -  & 0 & 0 \\
SO(4) & -&  - & 0 & - & - &  - & 0  & 0 & 0 \\
${\rm Z}_2\times $O(4) & -&  $\mu_1^2$ & 0 & - & $\lambda_1$ &  - & 0  & 0 & 0 \\
\col{SO(5)} & -&  $\mu_1^2$ & 0 & - & $\lambda_1$ & $2\lambda_1$ & 0 &
                                                                       0 & 0 \\ \hline 
\end{tabular}
\end{center}
\caption{Relations between the parameters of the $U(1)_Y$-invariant
  2HDM potential~\eqref{pot} for the 13 accidental
  symmetries~\cite{Pilaftsis:2011ed} 
in a diagonally reduced basis,
  where ${\rm Im} (\lambda_5) = 0$ and $\lambda_6=\lambda_7$. The symmetries in blue satisfy the natural alignment condition~\eqref{alcond}.}  
\label{tab1}
\end{table}

In terms of  the null-vector $R^A$, the  2HDM potential (\ref{pot}) 
takes on a simple quadratic form:
\begin{equation}
  \label{potR}
V\ =\ -\, \frac{1}{2}\, M_A\,R^A\: + \: \frac{1}{4}\, L_{AB}\, R^A R^B\; ,
\end{equation}
where $M_A$  and $L_{AB}$ are ${\rm SO}(1,5)$  constant `tensors' that
depend  on the  mass parameters  and quartic  couplings given in~\eqref{pot}   
 and    their   explicit    forms    may   be    found
in~\cite{Pilaftsis:2011ed,Maniatis:2007vn}.
Requiring   that    the   SU(2)$_L$   gauge-kinetic    term   of   the 
{\boldmath $\Phi$}-multiplet  remains canonical restricts  the allowed
set of  rotations from SO(1,5)  to SO(5),  where only  the  spatial components
$R^I$ (with $I=1,...,5$) transform and the zeroth component $R^0$
remains invariant. Consequently, in  the absence of  the hypercharge
gauge coupling and fermion Yukawa couplings, the maximal symmetry
group of the 2HDM is $G^R_{\rm  2HDM} = {\rm SO(5)}$. Including all proper, improper 
and semi-simple subgroups of SO(5), the accidental symmetries for the 2HDM 
potential were completely classified in~\cite{Battye:2011jj, Pilaftsis:2011ed}, as shown in Table~\ref{tab1}. Here we have used a diagonally reduced basis~\cite{Gunion:2005ja}, where ${\rm Im}(\lambda_5)=0$ and $\lambda_6=\lambda_7$, thus reducing the number of independent quartic couplings to seven. Each of the symmetries listed in Table~\ref{tab1} leads to certain constraints on the mass and/or coupling parameters.  From Table~\ref{tab1}, we find that 
there are {\it only} three symmetries, namely ${\rm Z}_2\times [{\rm O(2)}]^2$, O(3)$\times$ O(2), and  SO(5), which satisfy the natural alignment condition given by~\eqref{alcond}.\footnote{In Type-I 2HDM, there exists an additional possibility of realizing an exact Z$_2$ symmetry~\cite{Deshpande:1977rw} which leads to an exact alignment, i.e. in the context of the so-called inert 2HDM~\cite{Barbieri:2006dq}. In general, for $n$HDM with  $m<n$ inert scalar doublets, there are still three continuous alignment symmetries in the field space of the non-inert sector~\cite{Pilaftsis:2016erj}.} 
%Note that in all the three naturally aligned scenarios, $\tan\beta$ as given in (\ref{tanb}) `consistently' gives an {\em indefinite} answer 0/0. 
In the next two sections, we analyze each of these three realizations of the SM alignment. The simplest case based on  the SO(5) group has already been discussed in details in Refs.~\cite{Dev:2014yca, Dev:2015bta}. Our preliminary results reported for the other two cases are taken from our upcoming publication~\cite{next}. 
%%%%%%%%%%%%%%%%%%%%%%%%%%%%%%%%%%%%%%%%%%%%%%%%%%%%%%%%%%%%%%%%
\section{Maximally Symmetric 2HDM}\label{sec:4}
%%%%%%%%%%%%%%%%%%%%%%%%%%%%%%%%%%%%%%%%%%%%%%%%%%%%%%%%%%%%%%%
From Table~\ref{tab1}, we see that the maximal symmetry  group in the bilinear field space is SO(5), in which case the parameters of the 2HDM potential~\eqref{pot} satisfy the following relations:
\begin{align}
& \mu_1^2 \ = \ \mu_2^2\; , \quad m^2_{12} \ = \ 0\; , \quad \nonumber \\
& \lambda_2 \ = \ \lambda_1\; , \quad 
 \lambda_3  \ = \ 2\lambda_1\; , \quad 
\lambda_4 \ = \  {\rm Re}(\lambda_5) \  = \  \lambda_6 \ = \ \lambda_7\ =\ 0 \; ,
\label{so5}
\end{align} 
Thus, the 2HDM potential~\eqref{pot} is parameterized by just a {\em single} mass parameter $\mu_1^2=\mu_2^2\equiv \mu^2$ and a {\em single} quartic coupling $\lambda_1=\lambda_2=\lambda_3/2\equiv \lambda$, as in the SM: 
 \begin{align}
   \label{VSO5}
V \ & = \  -\,\mu^2\, \Big(|\Phi_1|^2+|\Phi_2|^2\Big)\: +\: \lambda\,
\Big(|\Phi_1|^2+|\Phi_2|^2\Big)^2 %\nonumber\\
 \  = \ -\: \frac{\mu^2}{2}\, {\bm \Phi}^\dagger\, {\bm \Phi}\ +\ 
\frac{\lambda}{4}\, \big( {\bm \Phi}^\dagger\, {\bm \Phi}\big)^2   \; .
\end{align}
%Note   that  the   MS-2HDM   scalar   potential
%in~(\ref{VSO5}) is  more minimal than the respective  potential of the
%MSSM at the  tree level. Even in the  custodial symmetric limit $g'\to
%0$, the latter possesses a smaller symmetry: ${\rm O}(2)\times
%{\rm  O}(3)  \subset  {\rm  SO}(5)$,  in  the  5-dimensional  bilinear
%$R^I$~space. 
%
%
Given the 
isomorphism of the Lie algebras  ${\rm SO(5)}  \sim  {\rm Sp}(4)$, the maximal symmetry group  
of the 2HDM in      the      original      {$\bm \Phi$}-field      space
is ${\rm G}^{\bm \Phi}_{\rm 2HDM} = \left[{\rm Sp}(4)/{\rm Z}_2\right] \times
{\rm SU(2)}_L$~\cite{Pilaftsis:2011ed, Dev:2014yca} in the custodial symmetry limit of vanishing  $g'$ and fermion Yukawa couplings. 
%We can generalize this result to deduce that in the custodial symmetry limit, the maximal symmetry group for an $n$ Higgs Doublet Model ($n$HDM) will be
%${\rm G}^{\bm \Phi}_{n{\rm HDM}} = \left[{\rm Sp}(2n)/Z_2\right] \times
%{\rm SU(2)}_L$.

%%%%%%%%%%%%%%%%%%%%%%%%%%%%%%%%%%%%%%%%%%%%%%%%%%%%%%%%%%%%%%%%%%%%
\subsection{Scalar Spectrum}\label{sec:spec-ms}
%%%%%%%%%%%%%%%%%%%%%%%%%%%%%%%%%%%%%%%%%%%%%%%%%%%%%%%%%%%%%%%%%%%%%%
Using the parameter relations given by \eqref{so5}, we find from~\eqref{mass1} and \eqref{mass3} that in the MS-2HDM, the CP-even scalar $H$ has mass $M_H^2=2\lambda_2
v^2$, whilst  the remaining four  scalar fields, denoted  hereafter as
$h$,  $a$ and  $h^\pm$, are  massless. This is a consequence of the Goldstone theorem, since after  electroweak symmetry  breaking, ${\rm  SO}(5) \  \xrightarrow{\langle \Phi_{1,2}\rangle \neq 0}  \ {\rm
  SO}(4)$. Thus, we identify
$H$ as  the SM-like Higgs  boson with the mixing  angle $\alpha=\beta$
[cf.~(\ref{HSM})], i.e. the SM alignment limit can be naturally attributed to the SO(5) symmetry of the theory.  

In the exact SO(5)-symmetric limit, the scalar spectrum of the MS-2HDM
is experimentally unacceptable. This is because the four massless pseudo-Goldstone
particles, viz.~$h$,  $a$ and $h^\pm$,  have sizeable couplings to  the 
SM $Z$ and $W^\pm$ bosons, and could induce additional decay channels,
such as~$Z\to  ha$ and $W^\pm  \to h^\pm h$, which  are experimentally
excluded~\cite{PDG}. However, as we will see below,
the   SO(5)  symmetry  may   be  violated
predominantly by  RG effects due to $g'$  and third-generation Yukawa
couplings, as well as by soft SO(5)-breaking mass parameters, thereby 
lifting the masses of these pseudo-Goldstone particles to be consistent with the experimental constraints.

%%%%%%%%%%%%%%%%%%%%%%%%%%%%%%%%%%%%%%%%%%%%%%%%%%%%%%%%%%%%%%%%%%%%%%%%%
\subsection{RG and Soft Breaking Effects \label{sec:RGE}}
\begin{figure}[t]
\centering
\includegraphics[width=6cm]{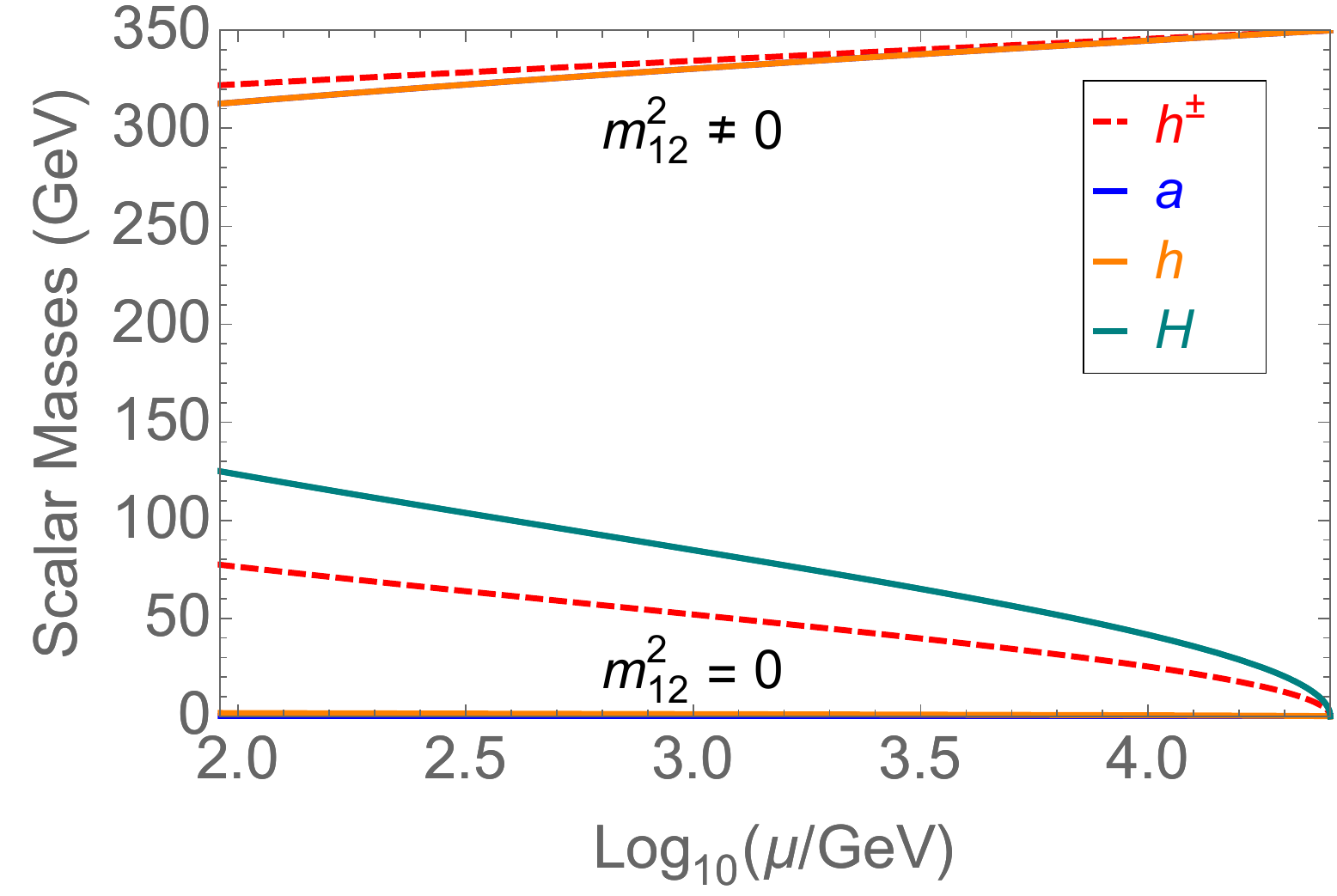}
%\hspace{0.5cm}
%\includegraphics[width=7cm]{fig1b.pdf}
\caption{The scalar spectrum  in the MS-2HDM  without and
  with soft-breaking effects.  
%induced by $m^2_{12}$. For $m^2_{12}=0$, the CP-odd scalar $a$ remains massless at tree-level, whereas $h$ and $h^\pm$ receive small masses due to the $g'$ and Yukawa coupling effects. For $m^2_{12}\neq 0$, one obtains a quasi-degenerate heavy Higgs spectrum, cf.~(\ref{mass-so5}). 
} \label{fig1}
%Here we have  chosen $\mu_X=2.5\times 10^{4}$ GeV,
%  $\lambda(\mu_X)=0$   and   $\tan\beta   =  50$   for   illustration.} 
\end{figure}

%%%%%%%%%%%%%%%%%%%%%%%%%%%%%%%%%%%%%%%%%%%%%%%%%%%%%%%%%%%%%%%%%%%
To calculate the RG and soft-breaking effects in a technically natural manner, we assume that the  SO(5)  symmetry is  realized  at  some  high scale~$\mu_X$ much above the electroweak scale. 
The
physical mass  spectrum at the  electroweak scale is then  obtained by
the RG evolution  of the 2HDM parameters given  by (\ref{pot}).  Using
the two-loop RG equations (RGEs) given in Ref.~\cite{Dev:2014yca}, 
we  first examine the
deviation of the Higgs spectrum  from the SO(5)-symmetric limit due to
$g'$   and  Yukawa   coupling   effects, in the absence of the soft-breaking term.   This   is  illustrated   in
Figure~\ref{fig1}  for a  typical choice  of parameters  in  the Type-II
realization of the  2HDM. We find that the RG-induced $g'$ effects only lift the charged Higgs mass
$M_{h^\pm}$, while the corresponding Yukawa coupling effects also lift
slightly  the  mass of  the  non-SM CP-even  pseudo-Goldstone
boson~$h$.  However, they still leave the CP-odd scalar $a$ massless,  which can be identified  as a
${\rm  U}(1)_{\rm PQ}$  axion~\cite{Peccei:1977hh}.  

Therefore, $g'$ and Yukawa coupling effects are
{\em  not} sufficient to  yield a  viable Higgs  spectrum at  the weak
scale, starting  from a  SO(5)-invariant boundary condition~\eqref{so5}  at some  high scale
$\mu_X$.   To minimally circumvent  this problem,  we include
soft SO(5)-breaking  effects, by assuming  a non-zero soft-breaking term ${\rm
  Re}(m_{12}^2)$.   
In the  SO(5)-symmetric limit  for the
scalar  quartic couplings,  but  with ${\rm  Re}(m_{12}^2)\neq 0$,  we
obtain the following mass spectrum [cf.~(\ref{mass1}) and \eqref{mass3}]:
\begin{eqnarray}
M_H^2 \ = \ 2\lambda_2 v^2\; , \qquad M_h^2 \ = \ M_a^2 \ = \ M^2_{h^\pm} \ = \
\frac{{\rm Re}(m^2_{12})}{s_\beta c_\beta} \; ,
\label{mass-so5}
\end{eqnarray}
as  well as  an equality  between  the CP-even  and CP-odd  mixing
angles: $\alpha = \beta$, thus predicting an {\it exact} alignment for
the  SM-like  Higgs  boson  $H$, simultaneously with  
an experimentally allowed heavy Higgs spectra (cf. Figure~\ref{fig1} for $m^2_{12}\neq 0$ case). Note that in the alignment limit, the heavy Higgs sector is exactly degenerate [cf.~(\ref{mass-so5})] at the SO(5) symmetry-breaking scale, and at the low-energy scale, this degeneracy is mildly broken by the RG effects. Thus, we obtain a {\em quasi-degenerate} heavy Higgs spectrum, which is a unique prediction of the MS-2HDM, valid even in the non-decoupling limit, and can be used to distinguish this model from other 2HDM scenarios. 

%%%%%%%%%%%%%%%%%%%%%%%%%%%%%%%%%%%%%%%%%%%%%%%%%%%%%%%%%%%%
\subsection{Misalignment Predictions \label{sec:MP}}
\begin{figure}[t!]
\centering
\includegraphics[width=6cm]{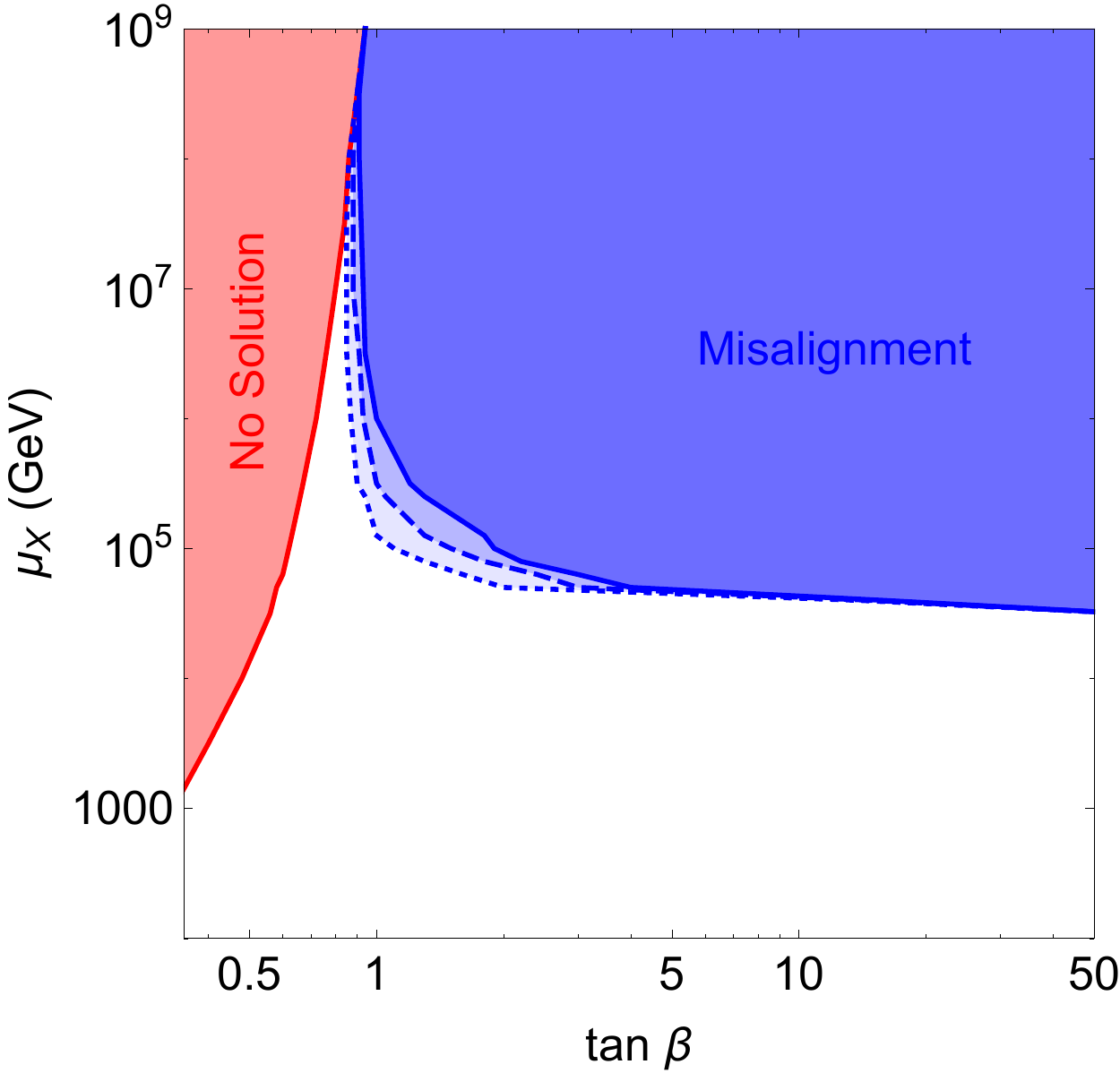}
\caption{The 
$1\sigma$ (dotted), $2\sigma$ (dashed) and $3\sigma$ (solid) exclusion contours (blue shaded region) 
from the alignment constraints in MS-2HDM. The red shaded region is theoretically excluded.%, as there is no solution to the RG equations up to two-loop order in this region.%, since the RGEs given in Appendix~\ref{app:RGE} do not have a solution 
%in this region. 
} \label{fig2} 
\end{figure}

%%%%%%%%%%%%%%%%%%%%%%%%%%%%%%%%%%%%%%%%%%%%%%%%%%%%%%%%%%%%
As discussed in Section~\ref{sec:RGE}, there will be some deviation
from  the  alignment  limit  in  the low-energy  Higgs  spectrum of the MS-2HDM due to RG and soft-breaking effects.   By requiring that the  mass and couplings of the  SM-like Higgs boson $H$  
are consistent with the LHC Higgs  data~\cite{Khachatryan:2016vau}, we derive predictions for the remaining
scalar spectrum and compare them with the existing (in)direct limits on the heavy
Higgs sector. 
%For the  SM-like Higgs boson mass,  we use  the $3\sigma$
%allowed range from the recent CMS and ATLAS
%Higgs    mass    measurements~\cite{coup, Aad:2014aba}: 
%$M_H \in  \big[124.1,~ 126.6\big]~{\rm GeV}$. 
We use the constraints in  the $(\tan\beta,~\beta-\alpha)$ plane derived from
a       recent       global       fit      for       the       Type-II
2HDM~\cite{Eberhardt:2013uba}, and require that for   a  given  set  of   SO(5)  boundary  conditions
$\big\{\mu_X,\tan\beta(\mu_X),\lambda(\mu_X)\big\}$, the
RG-evolved 2HDM  parameters at the  weak scale must satisfy  these 
alignment constraints  on the  lightest  CP-even Higgs  boson sector.   This puts stringent
constraints   on   the   MS-2HDM   parameter  space,   as   shown   in
Figure~\ref{fig2} by the blue shaded
region.    
In the red  shaded    region, there is no viable solution to the RGEs. 
We  ensure  that  the  remaining  allowed (white) region  satisfies  the
necessary theoretical constraints,  i.e.~positivity and vacuum  stability of the
Higgs     potential,    and     perturbativity     of    the     Higgs
self-couplings~\cite{review}.   From Figure~\ref{fig2},  we  find that
there  exists  an {\em upper}  limit  of  $\mu_X\lesssim  10^9$ GeV  on  the
SO(5)-breaking  scale   of  the   2HDM  potential,  beyond   which  an
ultraviolet completion  of the theory must be  invoked. 
% Moreover, for
%$10^5~{\rm GeV}\lesssim \mu_X \lesssim  10^9~{\rm GeV}$, only a narrow
%range of $\tan\beta$ values are allowed. 
The situation can be alleviated with the other two natural alignment scenarios listed in Table~\ref{tab1} and this will be the subject of the next section. 

\begin{figure}[t!]
\centering
\includegraphics[width=6cm]{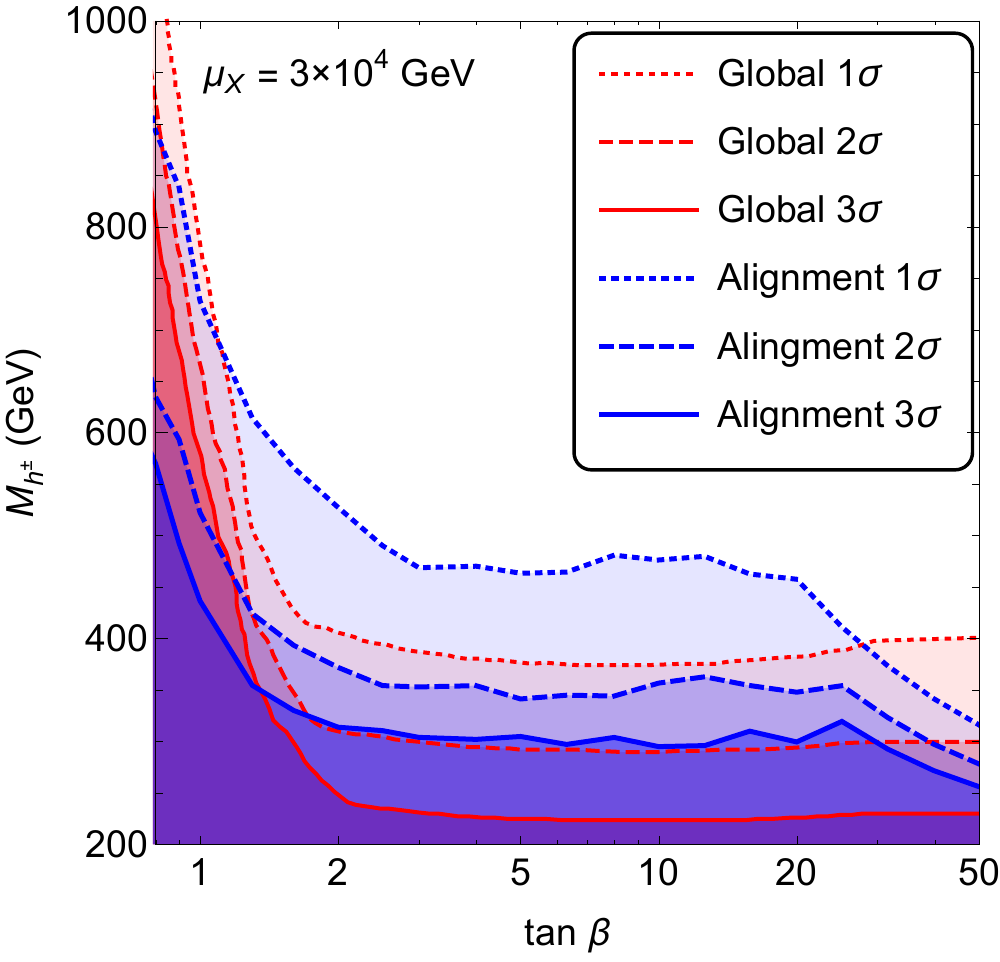}
\includegraphics[width=6cm]{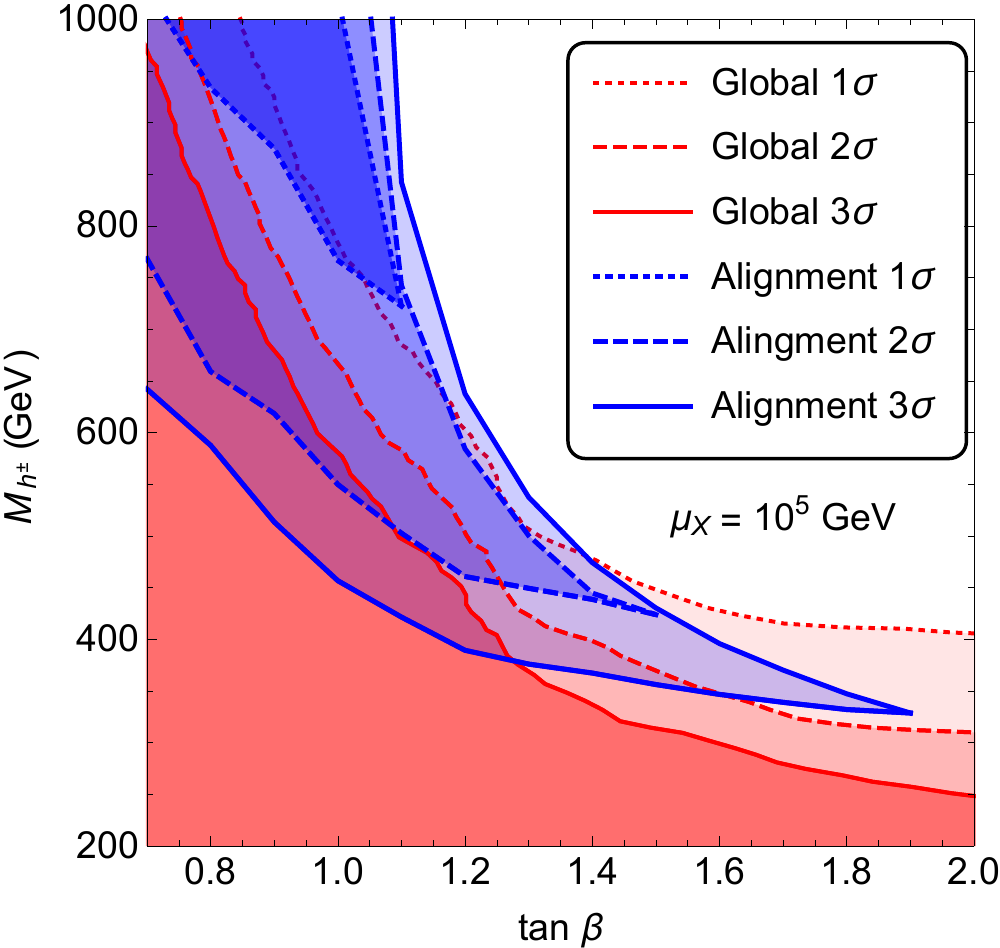}
\caption{{\em Left:} The $1\sigma$ (dotted), $2\sigma$ (dashed) and $3\sigma$ (solid) {\em lower} limits 
on the charged Higgs mass obtained from the alignment condition (blue lines) 
in the MS-2HDM with $\mu_X=3\times 10^{4}$ GeV. {\em Right:}  The $1\sigma$ (dotted), $2\sigma$ (dashed) and $3\sigma$ (solid) {\em allowed} regions from the alignment condition (blue lines) for  $\mu_X=10^{5}$ GeV. 
For comparison, the corresponding lower limits from a global fit 
are also shown (red lines). } \label{mhp}
\end{figure}

For  the  allowed   parameter  space  of  our  MS-2HDM   as  shown  in
Figure~\ref{fig2},  we obtain concrete  predictions for  the remaining
Higgs  spectrum.  In  particular,  the alignment  condition imposes  a
{\it lower} bound on the  soft breaking parameter Re$(m^2_{12})$, and hence,
on the heavy  Higgs spectrum. A  comparison of  the global fit limit on  the  charged  Higgs-boson mass  as  a function  of
$\tan\beta$~\cite{Eberhardt:2013uba} with our predicted limits from the alignment condition in the MS-2HDM for
a typical value of the boundary scale $\mu_X=3\times 10^{4}$ GeV is shown in Figure~\ref{mhp} (left panel).  It
is  clear that  the alignment  limits are  stronger than  the global fit 
limits, except  in the very small  and very large $\tan\beta$ regimes. 
%For 
%$\tan\beta\lesssim 1$ region, the indirect limit obtained  from 
%the $Z\to b\bar{b}$  precision observable becomes
%the strictest~\cite{Deschamps:2009rh, Eberhardt:2013uba}. Similarly, for the 
%large $\tan\beta\gtrsim 30$ case, the alignment limit can be easily obtained 
%[cf.~(\ref{chat})] 
%without requiring a large soft-breaking parameter $m_{12}^2$, and therefore, 
%the lower limit on the charged Higgs mass derived from the misalignment condition 
%becomes somewhat weaker in this regime. 

From Figure~\ref{fig2}, we note that for $\mu_X\gtrsim
10^5$~GeV, phenomenologically acceptable alignment is not possible in the MS-2HDM 
for large $\tan\beta$ {\it and} large $m^2_{12}$.  Therefore, we also get an   {\it  upper}  bound  on  the  charged
Higgs-boson mass  $M_{h^\pm}$ from the misalignment condition, depending on $\tan\beta$. This is illustrated in Figure~\ref{mhp} (right panel)  
for $\mu_X=10^5$ GeV.

Similar alignment constraints are obtained  for the heavy neutral pseudo-Goldstone
bosons $h$ and  $a$, which are predicted to be quasi-degenerate with the charged Higgs boson $h^\pm$ in the MS-2HDM [cf.~(\ref{mass-so5})]. The current experimental limits on the heavy neutral Higgs sector~\cite{PDG} are weaker than the alignment constraints in this case. Thus, the MS-2HDM scenario provides a natural reason for the absence of a heavy Higgs signal below the top-quark threshold, and this has important consequences for the heavy Higgs searches in the Run-II phase of the LHC, as discussed in Sec.~\ref{sec:5}.
%%%%%%%%%%%%%%%%%%%%%%%%%
\section{Planck-scale Perturbative Next-to-maximally Symmetric 2HDM} \label{sec:4b}
\begin{figure}[t!]
\centering
\includegraphics[width=6cm]{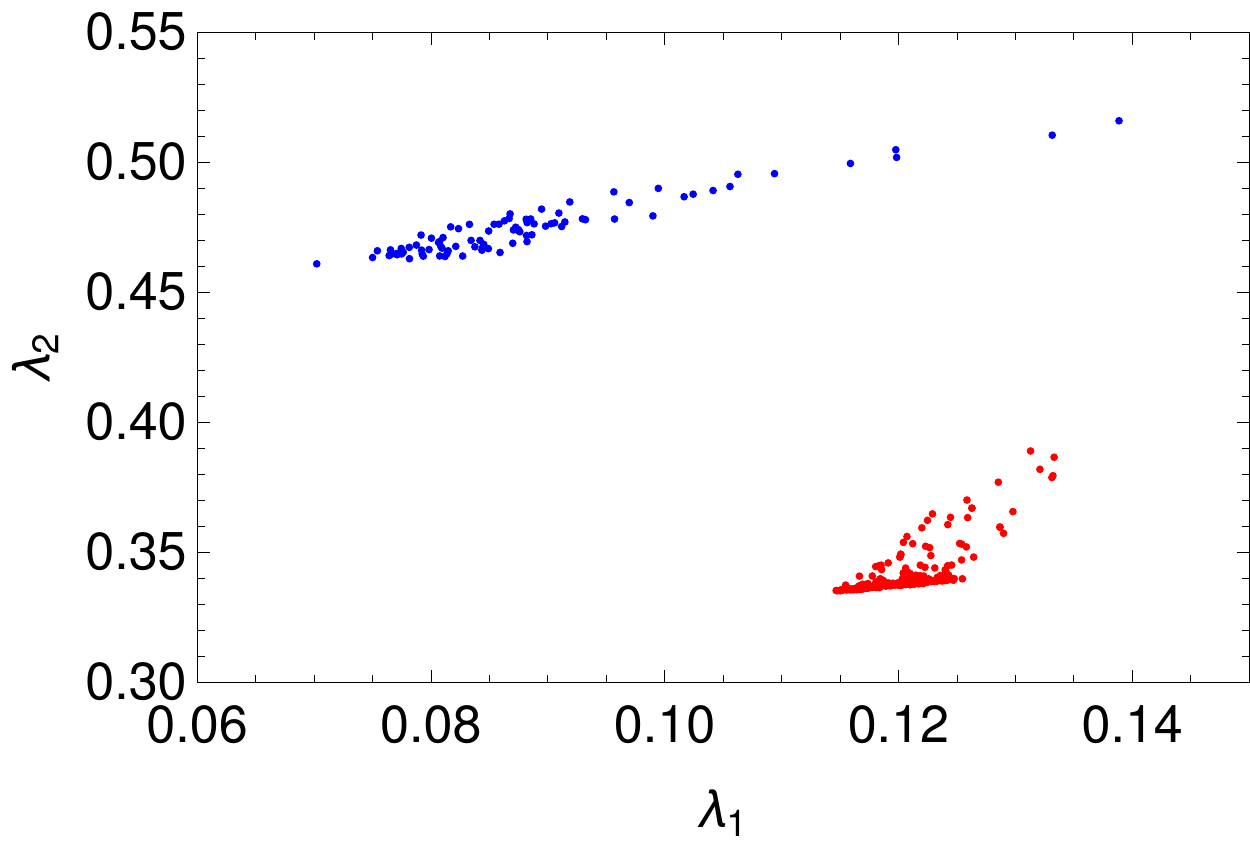}
\hspace{0.5cm}
\includegraphics[width=6cm]{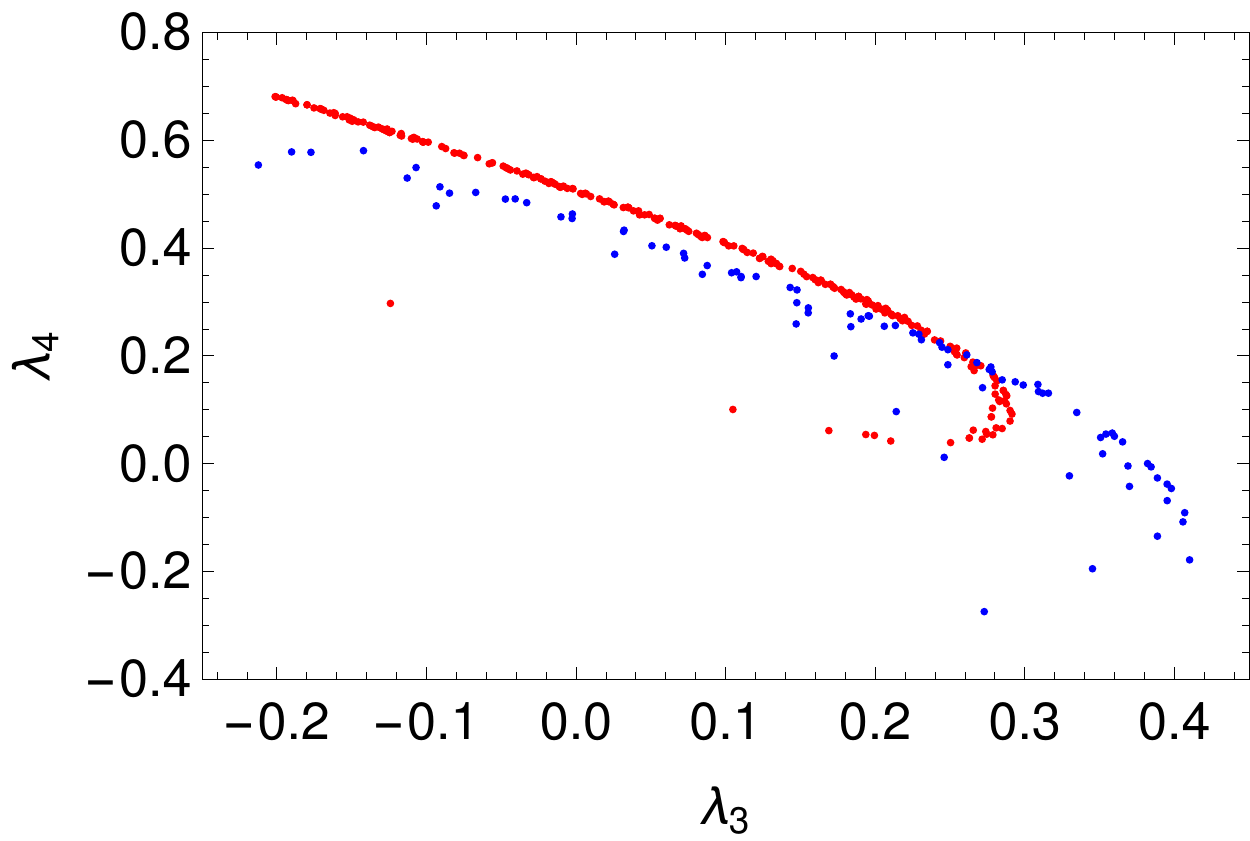} \\
\includegraphics[width=6cm]{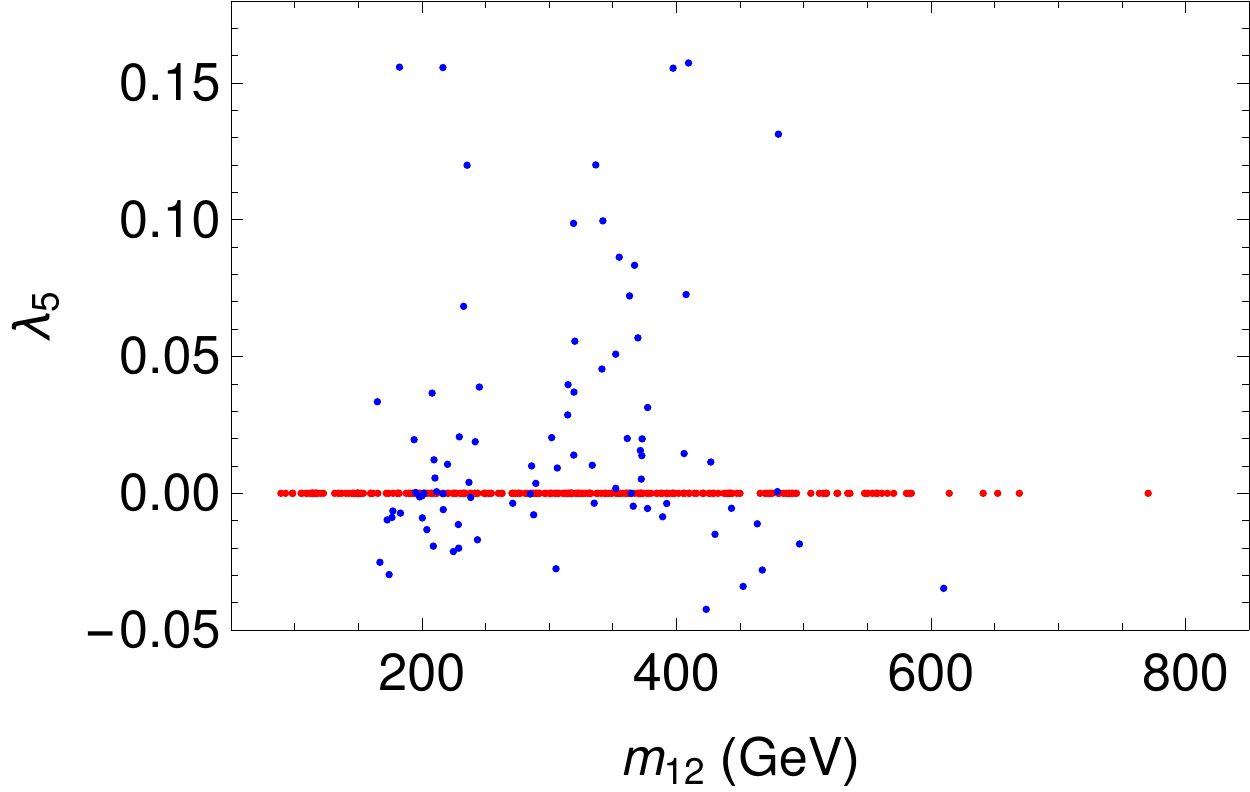}
\caption{The 2HDM parameter space at the EW scale satisfying the natural alignment condition up to the Planck scale. The red points are for the ${\rm O}(3)\times {\rm O}(2)$ case and the blue points are for the  ${\rm Z}_2\times [{\rm O}(2)]^2$ case. We have shown the results for the $\tan\beta=2$ case.}  \label{fig2a} 
\end{figure}

%%%%%%%%%%%%%%%%%%%%%%%%%%%%%%%%%%%%%
As discussed in Section~\label{sec:4}, the MS-2HDM scenario cannot be realized beyond $\mu_X\gtrsim 10^9$ GeV. However, as we show in this section, the other two natural alignment scenarios discussed in Table~\ref{tab1}, namely, those based on the ${\rm Z}_2\times [{\rm O}(2)]^2$ and ${\rm O}(3)\times{\rm O}(2)$ symmetry of the 2HDM potential, can be realized all the way up to the Planck scale, assuming no other intermediate new physics scale. To see this, we choose $\mu_X=M_{\rm Pl}=1.2\times 10^{19}$ GeV and apply the boundary conditions given in Table~\ref{tab1} for these two cases. For the ${\rm Z}_2\times [{\rm O}(2)]^2$  symmetry, this leaves three quartic couplings free, namely, $\lambda_{1,3,4}$, whereas for the ${\rm O}(3)\times {\rm O}(2)$ case, only two are free, namely, $\lambda_{1,3}$. We also allow for a non-zero $m_{12}^2$ to get a realistic scalar mass spectrum. Using the one-loop RGEs given in Ref.~\cite{Dev:2014yca}, we then study the top-down evolution of the 2HDM parameter space up to the weak scale and examine those points satisfying the vacuum stability, perturbativity and the LHC Higgs data, including the SM Higgs mass and the alignment constraints. Our numerical results are given in Figure~\ref{fig2a} for the $\lambda_1-\lambda_2$, $\lambda_3-\lambda_4$, and $\lambda_5-m_{12}$ parameter space at the weak scale, respectively. We have shown the results for the $\tan\beta=2$ case, and the corresponding results for higher $t_\beta$ values can be found in Ref.~\cite{next}.       

\begin{figure}[t!]
\centering
\includegraphics[width=6cm]{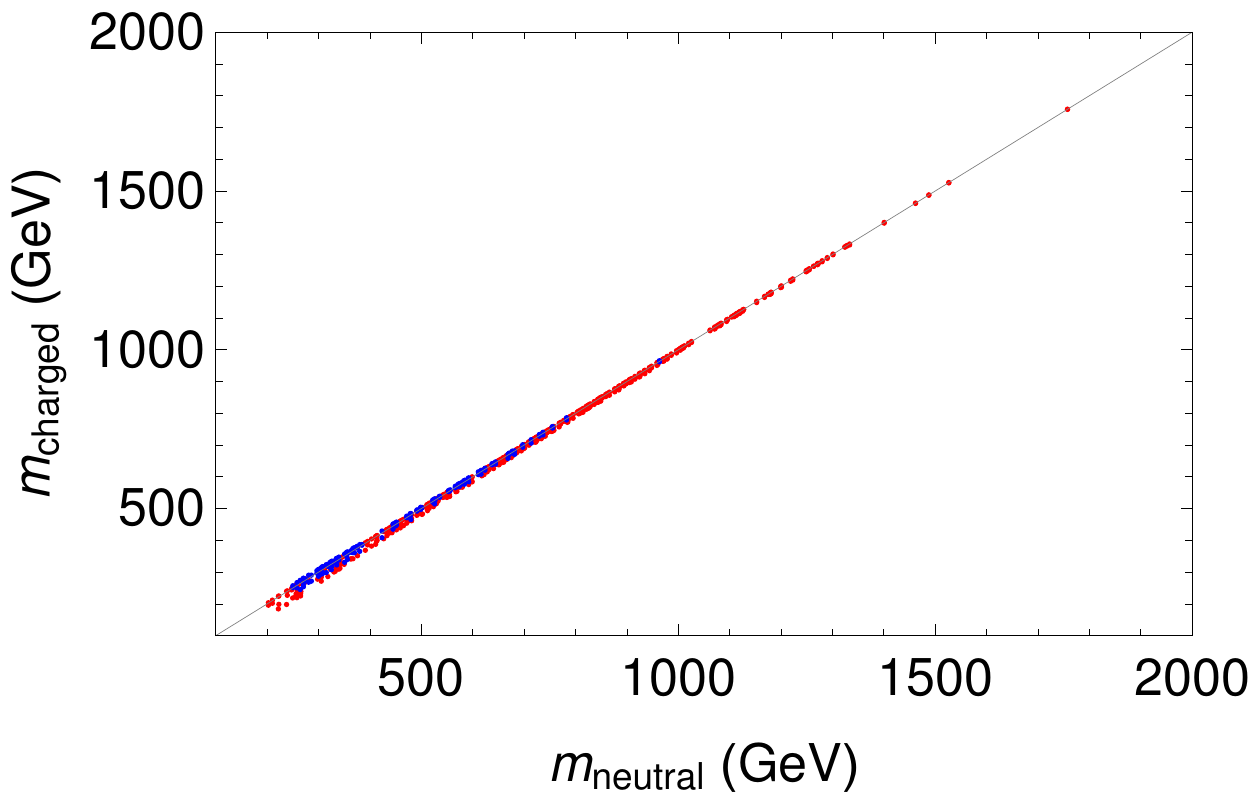}
\caption{The scalar mass spectrum at the EW scale satisfying the natural alignment condition up to the Planck scale. The red points are for the ${\rm O}(3)\times {\rm O}(2)$ case and the blue points are for the  ${\rm Z}_2\times [{\rm O}(2)]^2$ case. The gray line shows the exact degeneracy. We have shown the results for the $\tan\beta=2$ case.}  \label{fig2b} 
\end{figure}

The corresponding scalar mass spectra are shown in Figure~\ref{fig2b}, where in the $x$-axis, we have plotted the heavy neutral Higgs mass $m_h\simeq m_a$ and on the $y$-axis, the charged Higgs mass $m_{h^\pm}$. It is clear that as in the MS-2HDM case, the heavy Higgs spectrum is quasi-degenerate. Thus, we conclude that the near-degeneracy of the heavy Higgs spectrum is a generic prediction of the natural alignment scenario, irrespective of the underlying symmetry of the 2HDM potential. This is a key result that will be important for the collider signal analysis in the next section.

%%%%%%%%%%%%%%%%%%%%%%%%%%%%%%%%
\section{Collider Signatures in the Alignment Limit} \label{sec:5}

In the alignment limit, the  couplings of the lightest CP-even Higgs
boson are exactly  similar to the SM Higgs  couplings, while the heavy
CP-even    Higgs    boson  is gaugephobic.   Therefore,  two   of  the  relevant  Higgs
production mechanisms at the LHC,  namely, the vector boson fusion and
Higgsstrahlung processes are suppressed  for the heavy neutral Higgs sector. 
As a consequence, the only relevant production channels to probe the neutral Higgs
sector  of the  MS-2HDM  are the  gluon-gluon  fusion and  $t\bar{t}h$ 
($b\bar{b}h$) associated  production mechanisms at low (high) $\tan\beta$. 
For  the  charged Higgs sector of the MS-2HDM, the  dominant production 
mode is the associated production process: $gg\to \bar{t}bh^++t\bar{b}h^-$, irrespective of 
$\tan\beta$.

Similarly, for the decay modes of the heavy neutral Higgs bosons in the MS-2HDM, 
the $t\bar{t}$ ($b\bar{b}$) channel is the  dominant one for low (high) $\tan\beta$ values, 
whereas for the charged Higgs boson $h^{+(-)}$, the $t\bar{b}(\bar{t}b)$ mode is the 
dominant one for any $\tan\beta$. Thus, the heavy Higgs sector of the MS-2HDM can be effectively probed at the LHC through the final states involving third-generation quarks. 

%%%%%%%%%%%%%%%%%%%%%%%%%%%%%%%%
\subsection{Charged Higgs Signal} \label{sec5.2}
\begin{figure}[t]
\centering
\includegraphics[width=7cm]{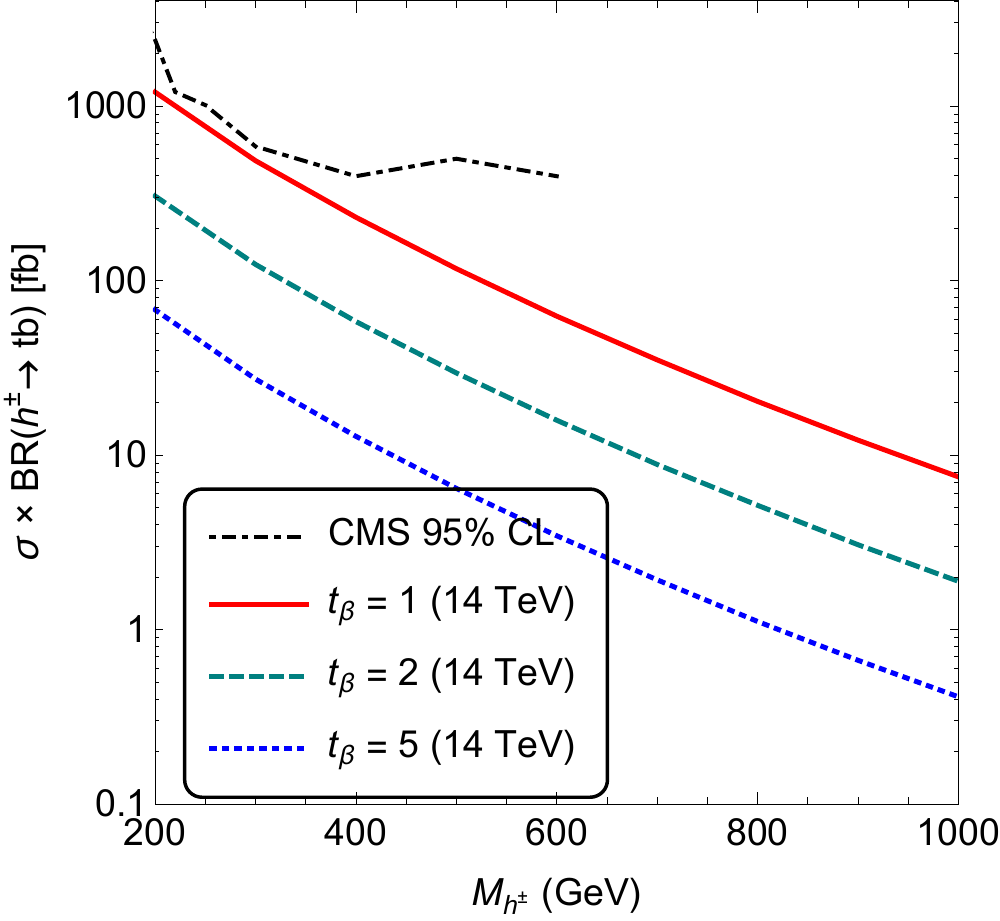}
\caption{Predictions for the cross section of the process (\ref{ttbb}) in the Type-II MS-2HDM 
at $\sqrt s=14$ TeV LHC for various values of $\tan\beta$. For comparison, we have also shown the current 95\% CL CMS upper limit from the $\sqrt s=8$ TeV data~\cite{tb}.}    
\label{ttbb-cross}
\end{figure}

%%%%%%%%%%%%%%%%%%%%%%%%%%%%%%%%
The most promising channel at the LHC for the charged Higgs boson in the MS-2HDM is 
\begin{equation}
gg\ \to\ \bar{t} b h^+ + t\bar{b}h^- \ \to\ t\bar{t} b \bar{b}\;.
\label{ttbb}
\end{equation}
Experimentally, this is a challenging mode due to 
large  QCD backgrounds and the non-trivial event topology, involving at least four 
$b$-jets~\cite{hwg}. Nevertheless, a recent CMS study~\cite{tb} has presented for the first time a realistic analysis of this process, in the leptonic decay mode of the $W$'s coming from top decays: 
\begin{equation}
gg\ \to\ h^\pm tb \ \to \ (\ell \nu_\ell bb)(\ell'\nu_{\ell'}b)b 
\label{ttbb-ll}
\end{equation}
($\ell,\ell'$ beings electrons or muons). Using the $\sqrt s=8$ TeV LHC data, they have derived 95\% CL upper limits on the production cross section $\sigma(gg\to h^\pm tb)$ times the branching ratio BR($h^\pm\to tb$) as a function of the charged Higgs mass, as shown in Figure~\ref{ttbb-cross}. In the same Figure, we show the corresponding predictions at $\sqrt s=14$ TeV LHC in the Type-II MS-2HDM for some representative values of $\tan\beta$. The cross section predictions were obtained at leading order (LO) by implementing the 2HDM in {\tt MadGraph5\_aMC@NLO}~\cite{mg5} and using the {\tt NNPDF2.3} PDF sets~\cite{nnpdf}.  A comparison of these cross sections with the CMS limit suggests that the run-II phase of the LHC might be able to probe the low $\tan\beta$ region of the MS-2HDM parameter space using the process (\ref{ttbb}).  Note that the production
cross  section  $\sigma(gg\to  \bar{t}  b  h^+)$ decreases  rapidly  with
increasing $\tan\beta$ due to the Yukawa coupling suppression, 
even though  BR($h^+\to
\bar{t}b$)  remains close to  100\%. Therefore, this channel is only effective for low $\tan\beta$ values. 

In order to make a rough estimate of the $\sqrt s=14$ TeV LHC sensitivity to the charged Higgs signal (\ref{ttbb}) in the MS-2HDM, we perform a parton level simulation of the signal and background events using {\tt MadGraph5}~\cite{mg5}. For the event reconstruction, we use 
some basic selection  cuts on  the transverse
momentum, pseudo-rapidity and dilepton invariant mass, following the CMS analysis~\cite{tb}:
\begin{eqnarray}
&& p_T^\ell \ > \ 20~{\rm GeV}, \quad |\eta^\ell| < 2.5,  \quad p_T^j \ > \ 30~{\rm GeV}, \quad |\eta^j| < 2.4, \quad \slashed{E}_T > 40~{\rm GeV}\nonumber \\
&& \Delta R^{\ell\ell} > 0.4, \quad \Delta R^{\ell j} > 0.4, \quad M_{\ell\ell} > 12~{\rm GeV}, \quad |M_{\ell\ell}-M_Z| > 10~{\rm GeV}.
\label{cut}
\end{eqnarray}
Jets are reconstructed using the anti-$k_T$ clustering algorithm~\cite{anti-kT} with a distance parameter of 0.5. Since four $b$-jets are expected in the final state, at least two $b$-tagged jets are required in the signal events, and we assume the $b$-tagging efficiency for each of them to be  70\%.

\begin{figure}[t]
\centering
\includegraphics[width=7cm]{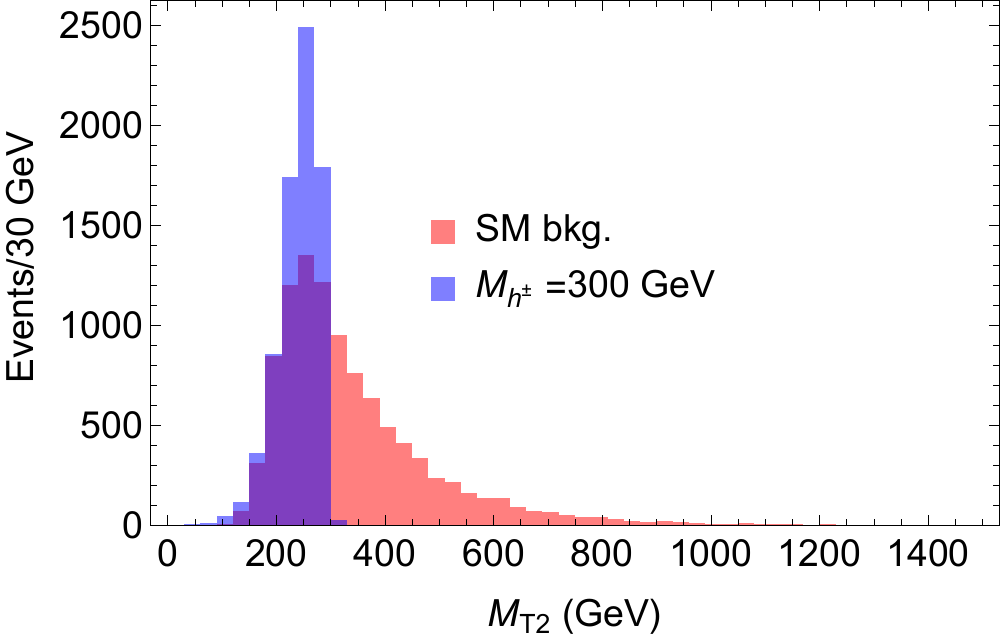}
\caption{An illustration of the charged Higgs boson mass reconstruction using the $M_{T2}$ variable. The irreducible SM background distribution is also shown for comparison.}    
\label{ttbb-dist}
\end{figure}

The inclusive SM  cross section for $pp\to t\bar{t}b\bar{b}+X$  is $\sim 18$ pb at NLO, with roughly 30\% uncertainty due to higher order QCD  corrections~\cite{pittau}. Most of the QCD background for the $4b+2\ell+\slashed{E}_T$ final state given by (\ref{ttbb-ll}) can be reduced significantly by reconstructing at least one top-quark. The remaining irreducible background due to SM $t\bar{t}b\bar{b}$ production can be suppressed with respect to the signal by reconstructing the charged Higgs boson mass, once a valid signal region is defined, e.g. in terms of an observed excess of events at the LHC in future.  For the semi-leptonic decay mode of top-quarks as in (\ref{ttbb-ll}), one cannot directly use an invariant mass observable to infer $M_{h^\pm}$, as both the neutrinos in the final state give rise to missing momentum. A useful quantity in this case is the $M_{T2}$ variable~\cite{mt2}, 
defined as 
\begin{eqnarray}
M_{T2} \ = \ \underset{\left\{ \slashed{\mathbf p}_{T_{\rm a}}+\slashed{\mathbf p}_{T_{\rm b}}=\slashed{\mathbf p}_T\right\}}{\rm min}\Big[{\rm max}\left\{m_{T_{\rm a}},m_{T_{\rm b}}\right\}\Big] \;,
\label{mt2}
\end{eqnarray}
 where $\{{\rm a}\}, \{\rm b\}$ stand for the two sets of particles in the final state, each containing a neutrino with part of the missing transverse momentum ($\slashed{\mathbf p}_{T_{\rm {a,b}}}$). Minimization over all possible sums of these two momenta gives the observed missing transverse momentum $\slashed{\mathbf p}_T$, whose magnitude is the same as $\slashed{E}_T$ in our specific case. In (\ref{mt2}), $m_{T_{i}}$ (with $i=$a,b) is the usual transverse mass variable for the system $\{i\}$, defined as 
\begin{eqnarray}
m_{T_{i}}^2 \ = \ \left(\sum_{\rm visible} E_{T_i}+\slashed{E}_{T_i} \right)^2- \left(\sum_{\rm visible} {\mathbf p}_{T_i}+\slashed{\mathbf p}_{T_i} \right)^2 \; .
\end{eqnarray}
For the correct combination of the final state particles in (\ref{ttbb-ll}), i.e. for $\{{\rm a}\}=(\ell \nu_\ell bb)$ and $\{{\rm b}\}=(\ell' \nu_{\ell'}bb)$ in (\ref{mt2}), the maximum value of 
$M_{T2}$ represents the charged Higgs boson mass, with the $M_{T2}$ distribution smoothly dropping to zero at this point. This is illustrated in Figure~\ref{ttbb-dist} for a typical choice of $M_{h^\pm}=300$ GeV. For comparison, we also show the $M_{T2}$ distribution for the SM background, which obviously does not have a sharp endpoint. Thus, for a given hypothesized signal region defined in terms of an excess due to $M_{h^\pm}$, we may impose an additional cut on $M_{T2}\leq M_{h^\pm}$ to enhance the signal (\ref{ttbb-ll}) over the irreducible SM background. 

\begin{figure}[t]
\centering
\includegraphics[width=7cm]{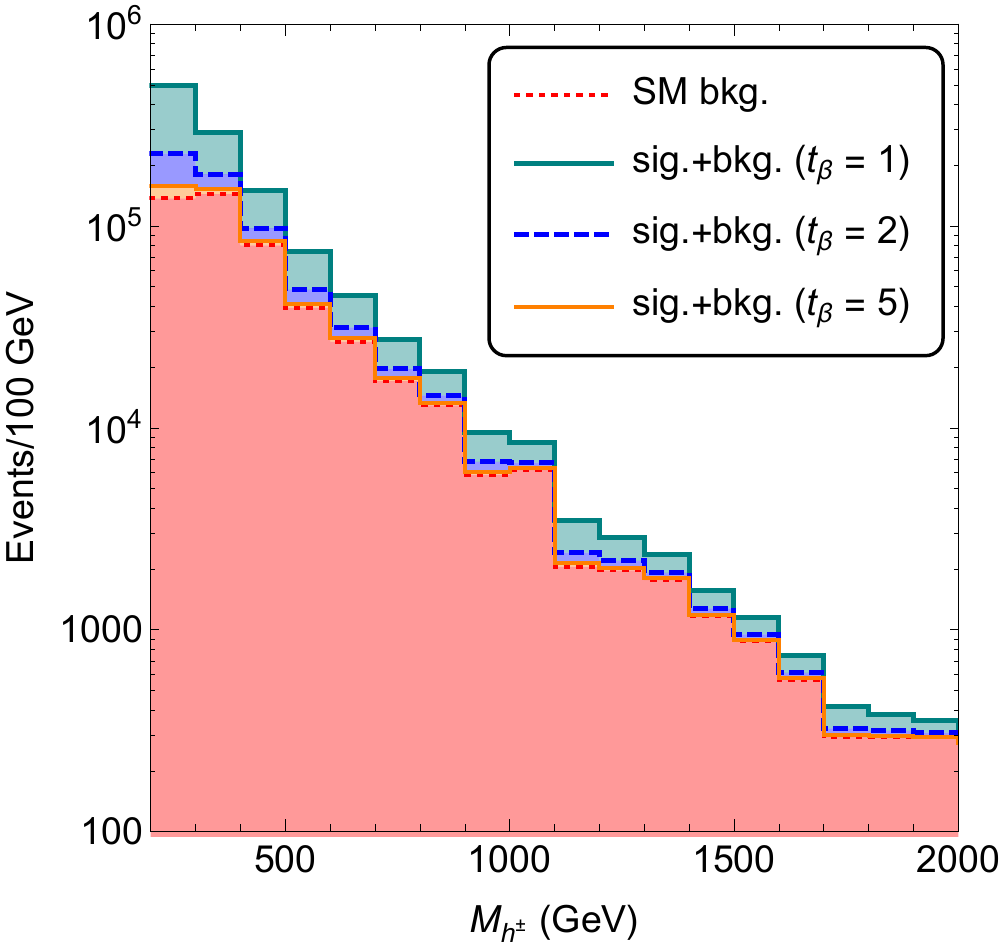}
\caption{Predicted number of events for the $t\bar{t}b\bar{b}$ signal
  from the charged pseudo-Goldstone boson in the MS-2HDM at $\sqrt
  s=14$ TeV LHC with $300~{\rm fb}^{-1}$ integrated luminosity. The irreducible SM background (red shaded) is controlled by assuming an efficient mass reconstruction technique~\cite{Dev:2014yca}. } 
\label{2tb}
\end{figure}

Assuming that the charged Higgs boson mass can be reconstructed efficiently, we present an estimate of the signal to background ratio for the charged Higgs signal given by (\ref{ttbb}) at $\sqrt s=14$ TeV LHC with 300 fb$^{-1}$ for some typical values of $\tan\beta$ in Figure~\ref{2tb}. Since the mass of the charged Higgs boson is a priori unknown, we vary the charged Higgs mass, and for each value of $M_{h^\pm}$, we assume that it can be reconstructed around its actual value within 30 GeV uncertainty.

\subsection{Heavy Neutral Higgs Signal} \label{sec5.3}

So far there have been no direct searches for heavy neutral Higgs bosons involving $t\bar{t}$ and/or $b\bar{b}$ final states, mainly due to the challenges associated with uncertainties in the jet energy scales and the combinatorics arising from complicated multiparticle final states in a busy QCD environment. Nevertheless, these channels become pronounced in the MS-2HDM scenario, and hence, we have made a preliminary attempt to study them in~\cite{Dev:2014yca} (see also~\cite{Gori:2016zto}). In particular, we focus on the search channel  
\begin{equation}
gg\ \to\  t\bar{t}h\  \to\ t\bar{t}t\bar{t}\; .
\label{4tp}
\end{equation}
Such four top final states have been proposed before in the context of
other  exotic  searches   at  the  LHC (see e.g.~\cite{4topother}).  However,  their relevance for heavy Higgs searches have not been explored so far. We note here that the existing 95\% CL experimental upper limit on the four top production cross section is 59 fb from ATLAS~\cite{atlas-4t} and 32 fb from CMS~\cite{cms-4t}, whereas the SM prediction for the inclusive cross section of the process $pp\to t\bar{t}t\bar{t}+X$ is about 10-15 fb~\cite{Bevilacqua:2012em}.

\begin{figure}[t!]
\centering
\includegraphics[width=8cm]{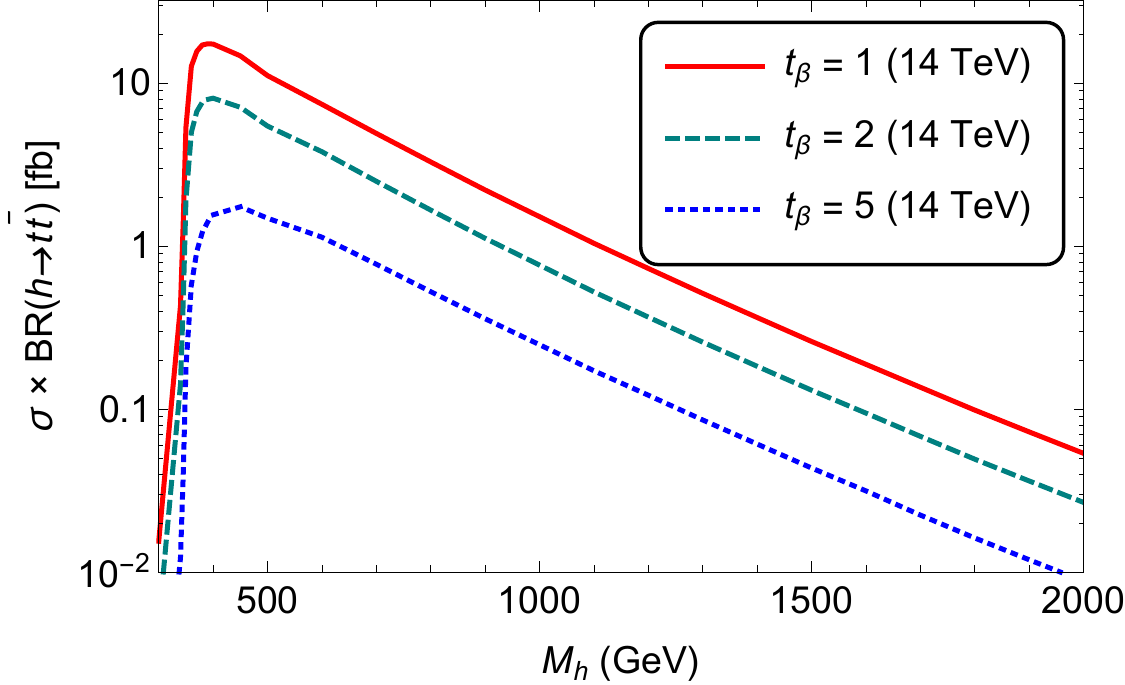}
\caption{Predictions for the cross section of the process (\ref{4tp}) in the Type-II MS-2HDM at $\sqrt s=14$ TeV LHC for various values of $\tan\beta$. } \label{4tx}
\end{figure}

To get a rough estimate of  the signal to background ratio for our 
four-top signal~\eqref{4tp},  we perform a  parton-level simulation of  the signal
and   background  events   at  LO  in   QCD   using  {\tt
  MadGraph5\_aMC@NLO}~\cite{mg5}   with   {\tt NNPDF2.3}  PDF sets~\cite{nnpdf}.  
For the  inclusive  SM  cross  section  for  the
four-top final  state at $\sqrt s=14$ TeV LHC, we obtain 11.85 fb, whereas  our proposed four-top 
signal  cross sections are
found   to  be   comparable  or   smaller  depending   on   $M_h$  and
$\tan\beta$, as shown in Figure~\ref{4tx}. However, since  we expect  one of  the  $t\bar{t}$ pairs
coming from  an on-shell  $h$ decay to  have an invariant  mass around
$M_h$,  we can  use this information to significantly boost the signal over the irreducible SM background. Note that all the predicted cross sections shown in Figure~\ref{4tx} are well below 
the current  experimental  upper bound~\cite{cms-4t}. 

Depending on the $W$ decay mode from $t\to Wb$, there are 35 final states for four top decays. According to a recent ATLAS analysis~\cite{thesis}, the experimentally favoured channel is the semi-leptonic/hadronic final state with two same-sign isolated leptons. Although the branching fraction for this topology (4.19\%) is smaller than most of the other channels, the presence of two same-sign leptons in the final state allows us to reduce the large QCD background substantially, including that due to the SM production of $t\bar{t}b\bar{b}+$jets~\cite{thesis}. Therefore, we will only consider the following decay chain in our preliminary analysis: 
\begin{eqnarray}
gg\ \to\  t\bar{t}h\  \to\ (t\bar{t})(t\bar{t}) \ \to \  
\Big( (\ell^\pm \nu_{\ell}b)(jjb)\Big)\Big((\ell'^\pm \nu_{\ell'}b)(jjb)\Big)\; .
\label{4t-ll}
\end{eqnarray}
For event reconstruction, we will use the same selection cuts as in (\ref{cut}), and in addition, following~\cite{thesis}, 
we require the scalar sum of the $p_T$ of all leptons and jets (defined as $H_T$) to exceed 350  GeV.    

\begin{figure}[t]
\centering
\includegraphics[width=8cm]{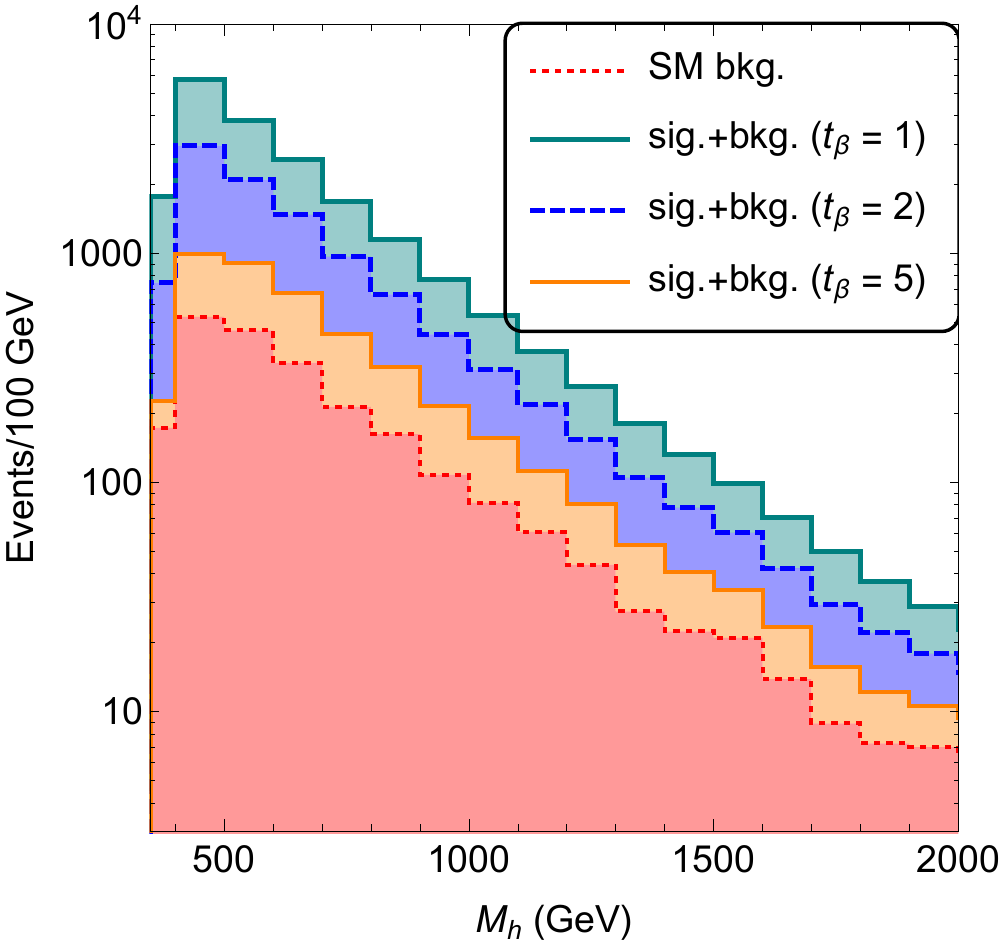}
\caption{Predicted number of  events for the $t\bar{t}t\bar{t}$ signal
  from  the neutral pseudo-Goldstone  boson in  the MS-2HDM  at $\sqrt
  s=14$ TeV  LHC with  $300~{\rm fb}^{-1}$ integrated  luminosity. The
  results are shown for three different values of $\tan\beta=$1 (green
  solid), 2 (blue  dashed) and 5 (orange solid).  The SM background (red dotted) is controlled by assuming an efficient mass reconstruction technique, as outlined in the text. } \label{4t}
\end{figure}

As in the charged Higgs boson case (cf.~Figure~\ref{ttbb-dist}), the heavy Higgs mass can be reconstructed from the signal given by (\ref{4t-ll}) using the $M_{T2}$ endpoint technique, and therefore, an additional selection cut on $M_{T2}\leq M_h$ can be used to enhance the signal over the irreducible background. 
Our simulation results  for the predicted number of signal and background events for the process (\ref{4t-ll}) at $\sqrt s=14$ TeV LHC with 300 fb$^{-1}$ luminosity are  shown in Figure~\ref{4t}. The signal events are shown for  three representative
values of $\tan\beta$. Here we vary the a priori unknown heavy Higgs mass, and for each value of $M_{h}$, we assume that it can be reconstructed around its actual value within 30 GeV uncertainty. From this preliminary
analysis, we  find that the
$t\bar{t}t\bar{t}$ channel provides the most promising collider signal
to  probe the  heavy Higgs  sector in  the MS-2HDM  at low  values of
$\tan\beta \lesssim  5$.

The above analysis  is also applicable for the  CP-odd Higgs boson
$a$,  which  has  similar  production  cross  sections  and  $t\bar{t}$
branching fractions as the CP-even Higgs $h$. However, 
the $t\bar{t}h(a)$ production cross section  as well as the $h(a)\to t\bar t$ branching
ratio decreases with  increasing $\tan\beta$. This is due  to the fact
that   the   $ht\bar{t}$   coupling   in  the   alignment   limit   is
$\cos\alpha/\sin\beta\sim \cot\beta$, which is same as the $at\bar{t}$
coupling. Thus,  the high $\tan\beta$ region of
the MS-2HDM  cannot be searched  via the $t\bar{t}t\bar{t}$ channel  proposed above,
and one  needs to consider  the channels involving  down-sector Yukawa
couplings, e.g. $b\bar{b}b\bar{b}$ and $b\bar{b}\tau^+\tau^-$~\cite{hwg}. It is also worth commenting here that the simpler process  $pp\to h/a \to t\bar{t}~(b\bar{b})$ at low (high) $\tan\beta$ suffers from a huge SM $t\bar{t}$ ($b\bar{b}$) QCD background, even after imposing an $M_{t\bar{t}~(b\bar{b})}$ cut. Some parton-level studies of this signal in the context of MSSM have been performed in~\cite{Djouadi:2013vqa}.

We should clarify that the results obtained in this section are valid only at the parton level. In a realistic detector environment, the sharp features of the signal [see e.g.,~Figure~\ref{ttbb-dist}] used to derive the sensitivity reach in  Figures~\ref{2tb} and \ref{4t}  may not survive, and therefore, the signal-to-background ratio might get somewhat reduced than that shown here. A detailed detector-level  analysis  of  these  signals,  including realistic  top reconstruction  efficiencies and smearing effects, as well as possible interference effects between the charged and neutral Higgs signals and with the SM background, is currently being  pursued in a separate dedicated study~\cite{next2}.

\section{Conclusions}\label{sec:6}

We provide a symmetry justification of the so-called SM alignment limit, independently of
the  heavy  Higgs  spectrum  and  the  value  of  $\tan\beta$ in the  2HDM.  
We show that in the Type-II 2HDM, there exist {\em only} three different symmetry realizations, which could lead to the SM alignment by satisfying the natural alignment condition~(\ref{alcond}) for {\em any} value of $\tan\beta$. In the context of the Maximally Symmetric 2HDM  based  on the  SO(5)  group,  we demonstrate how small
deviations from this alignment limit are naturally induced by RG 
effects  due  to  the   hypercharge  gauge  coupling  $g'$  and  third
generation Yukawa  couplings, which also break  the custodial symmetry
of  the theory.   In  addition, a  non-zero  soft SO(5)-breaking  mass
parameter is required to yield a viable Higgs spectrum consistent with
the  existing experimental constraints.   Employing the  current Higgs
signal strength  data from the  LHC, which disfavour  large deviations
from the alignment limit, we  derive important constraints on the 2HDM
parameter space.  In particular, we  predict lower limits on the heavy
Higgs spectrum,  which prevail the present  global fit limits in a  wide range of
parameter space.   Depending on the  scale where the  maximal symmetry
could be  realized in  nature, we  also obtain an  upper limit  on the
heavy Higgs masses in certain  cases, which could be probed
during  the  run-II phase  of  the LHC.   Finally,  we  have studied the collider signatures of the heavy Higgs sector in the alignment limit  beyond the top-quark threshold. We find that the final states involving third-generation quark final states can
become a valuable observational tool to directly probe the heavy Higgs
sector of the 2HDM in the alignment limit for low values of $\tan\beta$.

%\vfill\eject

\section*{Acknowledgements}
This work was  supported   by  the   Lancaster-Manchester-Sheffield
Consortium for Fundamental Physics under STFC grant ST/L000520/1. P.S.B.D. would like to acknowledge the local hospitality provided by the CFTP and IST, Lisbon where part of these proceedings was written.

\end{document}